\documentstyle[emulateapj,apjfonts,psfig]{article}

\lefthead{J.~M.~Miller et al.}  
\righthead{The Disk Wind in GRO~J1655$-$40}

\received{Date}
\revised{Date}
\accepted{Date}

\journalid{vol}{date}
\articleid{1}{4}
\paperid{id}

\cpright{AAS}{1999}
\ccc{x}

\begin{document}

\title{The Accretion Disk Wind in the Black Hole GRO~J1655$-$40}

\author{J.~M.~Miller\altaffilmark{1},
        J.~Raymond\altaffilmark{2},        
        C.~S.~Reynolds\altaffilmark{3},
        A.~C.~Fabian\altaffilmark{4}, 
        T.~R.~Kallman\altaffilmark{5},
        J.~Homan\altaffilmark{6}}

\altaffiltext{1}{Department of Astronomy, University of Michigan, 500
Church Street, Ann Arbor, MI 48109, jonmm@umich.edu}
\altaffiltext{2}{Harvard-Smithsonian Center for Astrophysics, 60
	Garden Street, Cambridge, MA 02138}
\altaffiltext{3}{Department of Astronomy, University of Maryland,
College Park, MD 20742}
\altaffiltext{4}{Institute of Astronomy, University of Cambridge,
Madingley Road, Cambridge CB3 OHA, UK}
\altaffiltext{5}{Laboratory for High Energy Astrophysics, NASA Goddard
Space Flight Center, Code 662, Greenbelt, MD 20771}
\altaffiltext{6}{MIT Kavli Institute for Astrophysics and Space
Resarch, 70 Vassar Street, Cambridge MA, 02139}


\authoremail{jonmm@umich.edu}

\label{firstpage}

\begin{abstract}
We report on simultaneous {\it Chandra}/HETGS and {\it RXTE}
observations of the transient stellar-mass black hole GRO~J1655$-$40,
made during its 2005 outburst.  {\it Chandra} reveals a line--rich
X-ray absorption spectrum consistent with a disk wind.  Prior modeling
of the spectrum suggested that the wind may be magnetically driven,
potentially providing insights into the nature of disk accretion onto
black holes.  In this paper, we present results obtained with new models for
this spectrum, generated using three independent photoionization
codes: XSTAR, Cloudy, and our own code.  Fits to the spectrum in
particular narrow wavelength ranges, in evenly spaced wavelength
slices, and across a broad wavelength band all strongly prefer a
combination of high density, high ionization, and small inner radius.
Indeed, the results obtained from all three codes require a wind that
originates more than 10 times closer to the black hole and carrying a
mass flux that is on the order of 1000 times higher than predicted by
thermal driving models.  If seminal work on thermally--driven
disk winds is robust, magnetic forces may play a role in driving the
disk wind in GRO~J1655$-$40.  However, even these modeling efforts
must be regarded as crude given the complexity of the spectra.  We
discuss these results in the context of accretion flows in black holes
and other compact objects.
\end{abstract}

\section{Introduction}
Accretion disks around black holes in X-ray binaries are sure to have
a role in nearly every aspect of such systems, from radiative energy
release to jet production.  Black hole disks can potentially even be
used to reveal black hole spin, Lense-Thirring precession, and other
aspects of General Relativity.  However, it is extremely difficult to
obtain clean measurements of the thermal continuum emission from black
hole disks.  Absorption along the line of sight, uncertainties in the
nature and degree of simultaneous hard X-ray emission, and
uncertainties in parameters like distance, inclination, mass accretion
rate, and the effects of radiative transfer through disk atmospheres
are just some of the difficulties associated with studies of the disk
continuum.  Fortunately, we can improve our understanding of black
holes and their accretion disks by focusing on timing studies,
emission lines broadened by orbital motions, and disk winds
(e.g. Strohmayer 2000, Miller 2007, Proga 2003, Miller et al.\ 2006) .
Of these, disk winds may seem to be the least telling, but winds can
in fact reveal a great deal about the nature of disks.

Accretion disks transfer matter radially inward while
transporting away angular momentum (for background, see Frank, King,
\& Raine 2002).  While theoretical work on this topic is quite
advanced, observational constraints are relatively poor, especially in
the case of black hole disks.  Internal magnetic viscosity may drive
disk accretion (e.g. Shakura \& Sunyaev 1973, Balbus \& Hawley 1991), and
could give rise to slow, dense winds originating close to the black
hole (e.g. Proga 2003).  An optically-thick blackbody-like spectrum
should also result from this process (Shakura \& Sunyaev 1973), though
this signature is a generic prediction of any internal viscosity
mechanism.  Alternatively, rigid magnetic fields may transport angular
momentum by allowing some matter to be accelerated along field lines
(Blandford \& Payne 1982; see also Spruit 1996).  The resulting winds may
be clumpy, and should show significant rotation.  A characteristic
thermal disk spectrum is not a requirement in such a scenario.

Clearly showing that either magnetic process is at work in accreting
black holes would shed important new light on accretion disk physics.
However, magnetic fields are not the only means of driving winds in
accreting systems, nor the most readily observable.  Radiation
pressure can effectively drive moderately ionized winds, due to spikes
in the cross section due to certain UV transitions (e.g. Cordova \& Mason
1982, Proga, Stone, \& Drew 1998).  Compton heating of the accretion
disk by a central source of X-rays can raise the local gas temperature
to the escape velocity, and drive thermal winds in accreting systems
(Begelman, McKee, \& Shields 1983, Woods 1996).  Differentiating
magnetic winds from radiatively and thermally-driven winds is largely
a question of accurately determining the launching radius, density,
and geometry of the wind.  Relative to radiatively and
thermally-driven winds, magnetic winds can originate closer to the
black hole and can have significantly higher density (e.g. Proga
2003).  Accurate modeling of sensitive high resolution spectra with
photoionization codes can determine these parameters with a fair degree
of confidence.

In FU Ori stars, spectra reveal clear evidence of magnetocentrifugal
winds (Calvet, Hartmann, \& Kenyon 1993), demonstrating that such
winds are important in at least some accreting systems.  Evidence for
magnetic winds from accretion disks around compact objects has been
lacking until recently.  Observations of one white dwarf system reveal
that radiative driving is insufficient, leading to the suggestion that
magnetic pressure may be important (Mauche \& Raymond 2000).  In at
least one Seyfert AGN, there is some indirect evidence for a
magnetocentrifugal wind (Kraemer et al.\ 2005).  A turning point may
have come with a {\it Chandra}/HETGS observation of the stellar-mass
black hole GRO~J1655$-$40 in outburst.  Modeling of that spectrum
suggested a slow, very dense wind originating close to the black hole.
The radius and density parameters and inferred geometry are all
suggestive of a wind driven by magnetic forces, and may signal a wind
driven by the Poynting flux generated by internal magnetic viscosity
(Miller et al.\ 2006; hereafter, Paper I).  It is only because all
other viable driving mechanisms appear to be ruled-out that magnetic
driving must be considered; we cannot measure Zeeman splitting in
X-rays to directly probe the magnetic field.  Applying the same logic,
the winds detected in the stellar-mass black hole candidates
H~1743$-$322 (Miller et al.\ 2006b) and 4U~1630$-$472 (Kubota et al.\
2007) may also be driven magnetically.  However, better--known driving
mechanisms are less rigidly excluded in these sources.

GRO~J1655$-$40 is a particularly interesting and well-observed black
hole X-ray binary.  Outbursts from this transient were previously
observed in 1994 and 1996.  The apparently superluminal jets observed
in radio band put GRO J1655$-$40 into the ``microquasar'' category
(see Hjellming \& Rupen 1995).  Of the approximately 20 X-ray binaries
in which dynamical constraints demand a black hole primary (see
McClintock \& Remillard 2006), the parameters of GRO J1655$-$40 are
best understood.  The binary is likely 3.2~kpc distant, and appears to
consist of a $7.0\pm 0.2~M_{\odot}$ black hole orbiting a
$2.3~M_{\odot}$ F3 IV -- F6 IV companion star in a 2.6-day orbit
(Orosz \& Bailyn 1996).  The binary is likely viewed at an inclination
of $67^{\circ}$, although the inner disk may be viewed at an
inclination as high as $85^{\circ}$ (Hjellming \& Rupen 1995).  An
alternative mass and distance have recently been suggested (Foellmi et
al.\ 2006); however, these values would not allow the companion star
to fill its Roche lobe (J. Orosz, private communication).  The
precision with which these parameters are known is of tremendous help
in efforts to estimate parameters like black hole spin, the mass
accretion rate, and other facets of the system and accretion flow.

Herein, we report on the results of more extensive modeling of the
disk wind spectrum of GRO~J1655$-$40.  In addition to simple
phenomenological modeling of the spectrum and order-of-magnitude
estimates regarding the nature of the wind, we have developed new
photoionization models.  The code described in Raymond (1993) and
Miller et al.\ (2006b) agrees well with the more widely used XSTAR and
Cloudy models.  Here, we use XSTAR and Cloudy because those codes
contain the most recent atomic rates and because they are thoroughly
documented and familiar to many astronomers.  Moreover, these codes
now allow users to develop grids of models that may be implemented
into fitting packages like XSPEC and ISIS, so that direct spectral
fitting and direct $\chi^{2}$ minimization techniques can be employed
to constrain the nature of complex spectra.  Netzer (2006) described
models for the wind in GRO~J1655$-$40 consistent with thermal driving,
and those parameters serve as an important point of reference in our
analysis.  Although some of the most extreme parameters reported in
Miller et al.\ (2006b) can be slightly relaxed, magnetic driving is
still required if present models of thermal driving are accurate.  The
inner radius at which the wind must be produced is at least an
magnitude smaller than is predicted by analytical and numerical
thermal wind models, and the mass outflow rate is at least three
orders of magnitude greater than predicted by the same models.

\section{Observations and Data Reduction}
{\it Chandra} observed GRO~J1655$-$40 on 2005 April 1, starting at
12:41:44 (UT, or MJD 53461.53); 44.6 ksec of net exposure time were
obtained after standard filtering.  The High Energy Transmission
Grating Spectrometer (HETGS) was used to disperse the incident flux
onto the ACIS-S array.  The array was operated in
``continuous-clocking'' because the nominal 2.8~ms frametime of this
mode prevents photon pile-up.  In order to limit telemetry saturation
and the loss of frames, we used a gray filter in the region around the
aimpoint to record only one in 10 incident zeroth order photons.
Recording some zeroth order photons is critical to establishing the
wavelength grid in later reduction.

The {\it Chandra} observation was processed using the CIAO reduction
and analysis suite, version 3.2.2.  The evt1 file was filtered to
accept only the standard event grades, to accept only events from the
nominal good-time intervals, and to reject events from bad pixels.
The ``destreak'' tool was run to remove the effects of streaking on
the ACIS CCDs.  Spectra were extracted using the default settings for
the tools ``tg\_resolve\_events'' and ``tgextract''.  We used canned
rmf files to produce arfs via the ``fullgarf'' tool.  For final
analysis, first-order HEG spectra and arfs, and first-order MEG
spectra and arfs, were added using the tool ``add\_grating\_spectra''.
(As any real spectral line within a band covered by multiple first-order
spectra must be present in {\it all} spectra, the individual first-order
spectra were frequently checked during the line identification process
to select only robust lines.)  The instrumental configuration and the
reduction procedure outlined above are fully consistent with prior
observations and analyses of observations made in this mode; for
additional discussion and context please see Miller et al.\ (2004,
2006b) and Cackett et al.\ (2007).

{\it RXTE} observed GRO~J1655$-$40 simultaneously with {\it Chandra},
to provide improved constraints on the broad-band X-ray continuum
emission and flux variability.  A 12.6~ksec {\it RXTE} observation was
obtained starting on 2005 April 1 at 13:47:12 (UT).  The tools and
packages in the HEASOFT version 6.04 suite were used to reduce and
analyze all {\it RXTE} data.  Of the independent proportional counter
units in the {\it RXTE}/PCA, PCU-2 is the best-calibrated.  We
therefore restricted our spectral reduction and analysis to this
detector.  Spectra from PCU-2 were extracted from data taken in
Standard-2 mode, providing full coverage of the 2.0--60.0~keV bandpass
in 129 channels every 16 seconds.  Data from all of the Xe gas layers
were combined.  The combined PCA spectra were corrected for detector
deadtime.  Background spectra were made using the tool ``pcabackest''
and the appropriate bright-source background files.  Redistribution
matrix files (rmfs) and ancillary response files (arfs) were made and
combined into a single file using the tool ``pcarsp''.  Fits to
spectra of the Crab reveal deviations from a simple power-law at the
0.6\% level; we therefore added 0.6\% systematic error to all PCA bins
using the tool ``grppha'' prior to spectral analysis (this is common
practice; e.g. Miller et al.\ 2006b).  Spectra from HEXTE-A cluster
were also reduced and analyzed. Spectra were made from standard
archive mode data, which has a nominal time resolution of 32 seconds
and covers the 10--250~keV band with 61 channels.  All spectra were
background--subtracted and deadtime--corrected using the standard
procedures.  We fit spectra from PCU-2 in the 3--25~keV band, and
spectra from HEXTE-A above 20~keV, as the instruments are well
calibrated in these ranges.

Spectra obtained with {\it RXTE} were analyzed using XSPEC version
11.3.2 (Arnaud 1996).  Joint fits to the simultaneous {\it RXTE} and
broad-band {\it Chandra} spectra were also made using XSPEC.  Detailed
analysis of {\it Chandra} spectra involving both phenomenological and
photoionization modeling of the absorption was done using ISIS version
1.3.3 (Houck \& Denicola 2000).  In the analysis that follows, errors
on continuum parameters are 90\% confidence errors, and errors on line
parameters and photoionization model parameters are 1$\sigma$ errors.

\section{Preliminary Analysis}

\subsection{The 2005 Outburst Profile}

{\it RXTE} monitored the 2005 outburst of GRO~J1655$-$40 at an
exceptionally high cadence, providing an excellent view of the source
behavior.  In this section, we briefly note some general points.  All
of the spectra, lightcurves, and power spectra from the large
monitoring campaign were obtained using the procedure outlined above
and the methods described in Homan (2005).  In order to plot the
source lightcurve in Crab units, the source count rate was converted
using the fact that the Crab gives 2250 counts per second per PCU.
This conversion is inexact because the Crab has a well-known power-law
spectrum, whereas black holes can have more complex spectra dominated
by thermal emission.  The hardness curve was produced by taking the
ratio of count rates in the 10--20~keV and 3--6~keV bands. Finally,
the rms variability curve was generated by making Fast Fourier
Transforms of each observation using the method detailed in Homan et
al. (2005), and obtain the fractional rms in the $128^{-1}--64$ Hz
band from the rms normalized and Poisson subtracted power spectra.

Figure 1 plots the resulting lightcurve, spectral hardness curve, and
fractional variability curve.  The {\it Chandra} and {\it RXTE}
observations analyzed in this work were obtained at the start of the
2005 outburst of GRO~J1655$-$40, during a complex rise phase spanning
roughly 100 days.  Black hole transients exhibit periods of rather
distinctive and correlated multi-wavelength phenomena; these are
sometimes called ``states'' (for a review, see the discussion of
states in McClintock \& Remillard 2006).  The ``high/soft'' or
``thermal dominant'' state is typified by low fractional variability
and a very soft (disk--dominated) spectrum.  The {\it Chandra} spectra
analyzed in this work were clearly obtained in the disk--dominated
``high/soft'' or ``thermal dominant'' state.

\subsection{Continuum Spectroscopy}

The observed continuum spectra are very soft.  This fact, coupled with
the low effective area of {\it Chandra} above approximately 8~keV,
makes it difficult to accurately characterize non-thermal components
that are either very steep or very weak using {\it Chandra} alone.
Conversely, {\it Chandra} is better suited for determinations of both
disk parameters and the equivalent neutral hydrogen column density
along the line of sight than {\it RXTE}.  We therefore determined the
continuum by using the strengths of each observatory.

A precise and detailed characterization of the continuum and any disk
reflection features (for a review, see Miller 2007) is beyond the scope
of this paper.  However, a robust determination of the continuum is
possible with simple fiducial models that allow the broad-band source
luminosity to be estimated accurately.  We fit all of the simultaneous
spectra with a phenomenological model consisting of disk blackbody
(Mitsuda et al.\ 1984) and power-law components, modified by
absorption in the ISM using the ``phabs'' model.  An overall constant
was allowed to float between the components to account for residual
calibration errors.  Fits to MEG spectra were restricted to the
0.65--5.0 keV (2.5--19\AA) band, and fits to HEG spectra were
restricted to the 1.2--10.0 keV (1.2--10\AA) band as the HETGS is best
calibrated for broad--band fitting in these windows.

We assumed the standard column density along this line of sight in all
fits ($N_{H} = 7.4\times 10^{21}~{\rm cm}^{-2}$, Dickey \& Lockman
1990).  Fits to the {\it RXTE} spectra suggest a disk--dominated
spectrum, with a power-law index of $\Gamma = 3.54(1)$.  In fits to
the {\it Chandra}/HETGS spectra, we then fixed this power-law index
and derived better constraints on the disk flux.  Fits with a this
power-law index and a disk component give an inner disk color
temperature of $kT = 1.35(1)$~keV.  The total model gives an
unabsorbed 0.65--10.0~keV flux of $4.0(1)\times 10^{-8}~ {\rm erg}~
{\rm cm}^{-2}~ {\rm s}^{-1}$ ($L_{X} = 5.0(1) \times 10^{37}~{\rm
erg}~ {\rm s}^{-1}$).  The errors are also fiducial because the
overall fit is poor ($\chi^{2}/\nu > 28$) owing to the strong
absorption lines in the {\it Chandra} bandpass and calibration
uncertainties.  However, the continuum spectrum is characterized very
well by these fits.  It is worth noting that the source is not clearly
detected above 30~keV.  This is consistent with the observation
occurring in a ``high/soft'' or ``thermal dominant'' state, as noted
above.

\section{Detailed Analysis and Modeling of the Disk Wind}
%
%
The absorption spectrum revealed in our observations of
GRO~J1655$-$40 is extremely complex.  This necessitates a careful,
step-wise approach to characterizing the spectrum and the properties
of the observed disk wind.  Both the analysis itself and its
presentation are geared toward organizing complex results from complex
models in the most clear and unbiased way possible.

First, we present a simple phenomenological description of the
absorption spectrum using independent Gaussian models for each of the
lines.  We then note some of the basic wind properties that can be
inferred even with such basic modeling.  Special attention is given to
some specific properties of the wind.  From there, we detail efforts
to model the wind using independent photoionization codes.  The
different codes are employed to mitigate biases that might be peculiar
to just one code.  To prevent biases related to wavelength selections
and fitting ranges, we made fits to specific wavelength regions, to a
broad wavelength band, and in evenly-spaced wavelength slices.  The
results of these steps are reported and discussed below.

\subsection{Phenomenological Line Modeling}
Closely following the procedure in Paper I, we made fits to the {\it
Chandra}/HETGS spectra in 2--4~\AA~ slices using fiducial local
continuum models and simple Gaussian functions to describe the
absorption lines.  The strengths of this procedure are that it is
model--independent and that it is useful for identifying lines,
measuring wavelength (velocity) shifts, and crudely estimating line
strengths.  We detect 90 absorption lines significant at the 5$\sigma$
level of confidence or higher.  A boot-strap approach of
searching for strong lines from abundant elements and then associated
weaker lines succeeds in identifying 76 of the lines.  The species
observed span 32 charge states.

The fits are shown in Figure 2 and Figure 3, and the line parameters
and identifications derived are given in Table 1 and Table 2.  The
lines were identified based on the wavelengths given in standard
spectroscopic references including Verner, Verner, \& Ferland (1996),
the NIST Atomic Spectra Database, and Nahar \& Pradhan (1999).
Oscillator strengths were also taken from these references.  The
column densities given in Table 1 and Table 2 were calculated assuming
that all lines were generated on the linear part of the curve of
growth (see Spitzer 1978).

The absorption lines detected show blue-shifts in the range of
300--1600~km/s, signaling an outflow.  No clear emission lines are
detected in the spectrum, strongly suggesting that the outflow does
not fill the entire volume available (see below).  The detection of Fe
XXII absorption lines at 11.770\AA~ and 11.920\AA~ is especially
fortuitous, as these lines serve as density diagnostics with a
sensitivity comparable to He-like line triplets (Mauche, Liedahl, \&
Fournier 2003).

There are a few major shortcomings of this simple phenomenological
approach.  First, some of the lines observed are likely to be
saturated.  This can be seen in the fact that trends in equivalent
width and column density for a given ion do not follow that expected
based on oscillator strength.  Gaussian line models are likely to be
inadequate when lines are saturated as this functional form misses
power in the wings of a line; a Voigt profile is the right functional
form in such circumstances.  For these reasons, the column densities
reported in Tables 1 and 2 should be regarded with caution.  For the
strongest lines expected from each charge state, this procedure can
even complicate measures of velocity widths and velocity shifts.  The
Fe XXV He-$\alpha$ line at 1.850\AA~ is not observed to have a
blue-shift, for instance, and this is likely an effect of saturation
making it difficult to accurately estimate velocities.  Finally, this
procedure is not guided by a self-consistent physical picture -- all
of the Gaussian models are independent and not governed by realistic
global plasma parameters.

\subsection{Basic Wind Properties}

The binary is likely viewed at an inclination of 67$^{\circ}$ (Orosz
\& Bailyn 1996).  Thus, our line of sight is already close to the
plane of the disk.  However, if we assume that the radio jet axis
coincides with the black hole spin axis, and that the inner disk is
anchored perpendicular to that axis (Bardeen \& Petterson 1975,
Blandford \& Znajek 1977), then we may view the {\it inner} disk at an
inclination of 85$^{\circ}$ (Hjellming \& Rupen 1995).  This might
suggest a viewing angle of only 5$^{ \circ}$ above the accretion disk,
but the need to see over the flared outer edge of the disk suggests a
lower limit of 6$^{\circ}$ (Vrtilek et al.\ 1990).  An upper limit on
the vertical extent of the gas of 12$^{\circ}$ can be derived from the
lack of emission features in the spectrum (see Paper I).

Thus, simple geometrical arguments suggest that the observed
absorption is an outflow along the plane of the disk, or at least an
outflow with a high density along the plane of the disk.  A hot
central corona of the kind sometimes invoked to account for hard X-ray
emission in black holes could not supply the range of charge states or
column densities observed.  Moreover, because the mass donor companion
star in GRO~J1655$-$40 is a low mass star (Orosz \& Bailyn 1996), the
observed wind must be expelled from material in the accretion flow.
An accretion disk wind is thus the only viable explanation for
the observed absorption spectrum.

In X-ray binaries with high mass companions, absorption can change
with the orbital phase of the binary as the viewing angle to the X-ray
source through the companion wind changes.  Such a variation is not
expected in a low-mass X-ray binary like GRO~J1655$-$40.  Indeed,
when the observation is divided into time segments, the resulting
spectra show that the lines observed are effectively constant across
the entire observation.  Similarly, no significant variability is seen
when spectra selected from intervals of low and high count rates are
compared.  These findings signal that the wind is not likely due to
irradiation of a phase-dependent structure.  Varying degrees of
absorption were seen in independent observations with {\it
XMM-Newton}, but those variations were driven by changes in the X-ray
state of the source and dramatic changes in the ionizing flux (Diaz
Trigo et al.\ 2007, Sala et al.\ 2007).

More importantly, the lack of strong variability in the spectrum
suggests that partial covering of the central engine along the line of
sight is unlikely.  Partial covering can occur when an absorber has
significant and variable density structure.  The density structures
must be small compared to the central engine, and so ``clumps'' of
dense material must be invoked (the winds of massive stars are known
to be clumpy, for instance).  If the wind in GRO~J1655$-$40 was
concentrated into clumps, they would travel on the order of
$10^{12}$~cm during our observation, assuming the lowest values of the
blue-shifts observed (300~km/s).  That distance is of the same order
of magnitude as the size of binary system itself.  If the wind were
clumpy, blobs would easily move into and out of our line of sight
during the observation, strongly affecting the degree of absorption
observed.  The absence of strong variability makes partial covering
unlikely in GRO~J1655$-$40.  

\subsection{Differentiating Wind Driving Mechanisms}

Under the right conditions, radiation pressure can efficiently drive
winds.  At moderate ionization parameters, the cross section for
certain UV line transitions greatly increases, and allows momentum to
be transferred to the wind effectively.  A strong source of UV
radiation and moderately ionized gas are the only real requirements.
Thus, radiation pressure is the means by which winds are driven from O
stars, and it is undoubtedly important in driving some winds from
accreting white dwarfs and AGN.  However, for ionization parameters of
$\xi = 10^{3}$ and higher, radiation pressure supplies little
additional driving force (e.g. Proga, Stone, \& Kallman 2000).  The
fact that the UV zone of the disk is not the most central zone in
X-ray binaries also makes it difficult to drive winds through line
force unless the wind density is extremely high (Proga \& Kallman 2002).

A strong majority of the lines observed in GRO~J1655$-$40 are from
He-like and H-like charge states; there are very few L shell
transitions and very few UV transitions remaining.  Even from the
crude analysis detailed above, then, it is clear that radiation
pressure cannot drive the wind observed.  The detailed photoionization
modeling outlined below further confirms this finding.

Thermal winds can originate when the (flared) outer accretion disk is
irradiated by the central X-ray source.  If Compton heating of the
disk surface is able to raise the gas temperature to that
corresponding to the local escape speed, the heated material can flow
out of the system.  Both analytic (Begelman, McKee \& Shields 1983; BMS)
and numerical (Woods et al. 1996) models of thermally driven winds
from X-ray illuminated accretion disks have been developed.  The numerical
models eliminate the most serious approximations of BMS, namely the
simplified flow geometry and the neglect of the photoionization
heating and collisional excitation cooling that dominate in the gas
that produces the observed absorption lines.  Nevertheless, the
numerical models agree quite well with the analytic ones.

The models predict that a thermally driven wind can arise outside 0.1
$R_{C}$ (BMS) or 0.2 $R_{C}$ (Woods et al. 1996).  The Compton
radius is given by $R_{C} = (1.0\times 10^{10})\times
(M_{BH}/M_{\odot})/T_{C8}$~ (where $T_{C8}$ is the Compton temperature
in units of $10^8$~K).  In Paper I, we used the gas temperature in
this equation; using the Compton temperature derived in the
photoionization models described below, we find $T_{C} \simeq 1.4
\times 10^{7}$~K and $R_{C} \simeq 10^{11.7}$~cm, in agreement with
an independent analysis by Netzer (2006).

Even if the gas were beyond 0.1 $R_{C}$, the density predicted by the
thermal wind models is far lower than observed.  Equation 4.8 of Woods
et al. (1996), or their Figure 16 scaled from a $10^{8}~M_{\odot}$
black hole to a 7 $M_{\odot}$ black hole, predict the peak mass loss
rate per unit area.  For an ionizing spectrum that gives a Compton
temperature close to that found in our models or Netzer's, the maximum
is $6 \times 10^{-6} \rm g~cm^{-2}~s^{-1}$, compared with an observed
mass flux of about $10^{-2} \rm g~cm^{-2}~s^{-1}$.  Woods et al. find
that the flow is nearly vertical between 0.1 $R_{C}$ and $R_{C}$,
and the high inclination angle means that the actual speed is
considerably larger than the line-of-sight component, so the predicted
density is still smaller.

If a wind originates significantly closer to the black hole than is
predicted by thermal driving models, and/or if the density observed is
significantly above the predictions of the same models, magnetic
driving is the only viable means of powering a disk wind.  Density can
be deduced from density-sensitive line ratios, and from
photoionization modeling to a lesser extent.  The inner radius at
which absorption is observed to occur can be derived from
photoionization modeling.  Differentiating magnetic and thermal
driving mechanisms, then, is largely a question of accurately
determining the radius and density of the wind observed.

\subsection{On the density of the wind}

The brief discussion above makes it clear that accurately estimating
the density of a disk wind is critical to understanding its nature.
In emission line spectra, density is often constrained using He-like
triplet lines, but the observed spectrum lacks both emission lines and
clear evidence of intercombination and forbidden lines.  However, Fe
XXII absorption lines {\it are} present in the spectrum (see Figure 3,
and Table 2).  The relative populations of the fine structure states
of the ground level of Fe XXII, $2p~^2P_{1/2}$ and $^2P_{3/2}$,
provide a density diagnostic at $n_e~\sim 10^{12}-10^{15}~\rm cm^{-3}$
(Mauche et al.\ 2003).

We have measured the equivalent widths of the Fe XXII lines using a
common continuum for both lines and adding the constraint that the
separation equal the known separation of these lines.  This
methodology is slightly revised and improved from Paper I, and it
provides a tighter constraint on the line ratio.  The new equivalent
widths are 16.0$\pm$0.6 and 12.2$\pm$0.7 m\AA\/ for the 11.77 and
11.92 \AA~ lines, respectively.  Taking the oscillator strengths and
the effects of slight saturation into account, this implies 0.62 $\le$
$n_2/n_1$ $\le$ 0.76, where $n_2$ and $n_1$ are the populations of the
J=3/2 and 1/2 states, respectively.  Figure 4 shows the population
ratio computed with CHIANTI (Landi et al. 2006) for temperatures
between $10^5$ and $10^{6.3}$ K, corresponding to the upper and lower
stable temperature branches of the photoionization equilibrium.  We
note that radiative excitation of permitted lines could in principle
increase $n_2/n_1$.  However, even if the very steep power law extends
to wavelengths near 100~\AA, radiative excitation only makes a
significant difference if the gas is located within about $10^9$ cm of
the black hole.

In Paper I, we considered both the high and low temperature limits to
be acceptable possibilities, but it seems likely that if the Fe XXII
were formed in the low temperature branch, other lower Fe ions would
also be present.  Therefore, it appears that the $1 - 2 \times 10^6$ K
range of the upper branch as predicted by our models is appropriate
(this was confirmed independently using CLOUDY and XSTAR), and the
density is near to or within the range 13.6 $\le$ log($N_e$) $\le$
13.8.  It is reasonable to expect that the most ionized part of the
wind originates close to the black hole, and indeed most of the lines
observd are from He-like and H-like charge states.  If there is any
fall in density with radius, then the density range quoted here might
be regarded as an effective lower bound.

In contrast, Netzer (2006) presents two closely-related models for the
disk wind in GRO~J1655$-$40 that are consistent with thermal driving.
The density range found above is 3--6 times the densities at the inner
edges of the models described by Netzer (2006).  The densities of
those models decrease with radius, so the average densities are around
half the densities at the inner edge.  The density distributions for
both models predict column densities $N_2/N_1$ close to 0.1, a factor
of 7 below the observed value.

The models developed by Netzer (2006) are partially motivated by a
different interpretation of the Fe XXII lines.  A brief comparison to
those models provides a useful mechanism for discussing the Fe XXII
lines and density determination in more detail.  The Netzer models
assume that the 11.77 \AA~ line is saturated, and that the absorbing
gas covers only 75\% of the X-ray emitting region.  To support this
interpretation, Netzer suggests that two unidentified lines at 11.54
and 11.42 \AA~ are Fe XXII lines with smaller oscillator strengths.
With wavelengths from CHIANTI (Young et al. 2003) it is more logical
to identify the line at 11.47 \AA~ with a blue-shifted Fe XXII line at
11.51 \AA.  In that case, the equivalent widths of these lines from
the lower fine structure level but with small oscillator strengths
would be comparable to that of the 11.92 \AA~ line, and this is more
or less what is observed (see Table 2).

The latter interpretation cannot be correct, however, because it also
implies that the $2s^22p~^2P_{1/2}~-~2s^24d~^2D_{3/2}$ line at 8.97
\AA~ should also be stronger then the 11.51 \AA~ line.  An
unidentified line at 8.96 \AA~ is listed in Paper I with an equivalent
width of 2.6 m\AA~ (also see Table 2).  This is exactly what is
predicted given the oscillator strengths from CHIANTI (Young et
al. 2003) and the column density derived from the 11.77 \AA~ line
under the assumption that both are {\it optically thin}.  It is at
least a factor of 4 smaller than the value expected from Netzer's
suggestion of saturation and a partial covering factor.  It is likely
that the 8.96 \AA~ line is Fe XXII and that the features near 11.41
and 11.54 \AA~ are dominated by unidentified lines.  A broad feature
near 8.7 \AA~ is consistent with the Fe XXII excitations to the 2s2p4d
$^2P_{1/2}$ and $^2D_{3/2}$ levels, again under the assumption that
the 11.77 \AA~ line is unsaturated.

The analysis above strongly argues that the best description of the
spectrum and most physically plausible scenario is that the density
sensitive Fe XXII lines are optically thin and that the covering
factor along the line of sight is near to unity.  Interpreting the
11.77 \AA~ line as heavily saturated is inconsistent with the data.
As a result, the thermal driving models presented by Netzer (2006) use
a density that is too low by a factor of approximately 5--10.

\subsection{Direct Photoionization Modeling}

The photoionization models presented in Paper I (and Netzer 2006) were
matched to the equivalent widths of the absorption lines, but they did
not employ direct spectral fitting.  This is a standard procedure for
the application of photoionization codes to observations of a modest
number of absorption lines.  It has the advantage that the user can
focus on what he or she considers to be the key absorption lines,
ignoring lines that provide only redundant information, and it makes
it possible to take thei assumptions about uncertainties in atomic
rates into account.  (It also forces the user to understand which
features of the data the models are trying to match in achieving a
best fit.)  Among the disadvantages of this procedure are that it can
be slow and subjective, and that it makes error estimation difficult.

More robust estimates of line parameters and saturation in X-ray
spectra require proper direct spectral fitting.  Spectrometers like
the HETGS scatter light and can have broad line response functions, so
lines that appear not to be saturated might actually be black.  For
instance, Figures 5 and 6 show fits to the Fe K band with XSTAR and
CLOUDY (more details are given below).  The Fe XXV and XXVI lines at
1.8510(4) and 1.7714(5)\AA~ do not appear to be black.  The best fit
Cloudy model for the Fe K band is reproduced in Figure 7 {\it prior}
to convolving the model with the HETGS response function.  Clearly,
the Fe XXV and Fe XXVI lines are actually black.  This is important
because it means that partial covering is not required by the data.

XSTAR (Kallman \& Bautista 2001) and Cloudy (Ferland et al.\ 1998) are
the best known, best tested and best documented general purpose
photoionization codes presently available.  In addition to generating
new models using our own code, we have also generated a number of
models using XSTAR and CLOUDY.  This step is taken to add context, to
test the results presented in Paper I, and to assess how
model--dependent any particular results may be.  Importantly, new
functionality has been added to both XSTAR and Cloudy that allows
grids of models to be generated and implemented into XSPEC and ISIS
for direct spectral fitting.  This functionality makes it possible to
obtain the best possible constraints, and to obtain errors through
direct comparison to data and minimization of the $\chi^{2}$
goodness-of-fit statistic.

The code used in Paper I, XSTAR, and Cloudy differ from each other in
minor ways, but all meet the Lexington benchmarks (Ferland 1995).
Unlike the crude line-by-line fits made above, all three of these
models describe lines using Voigt profiles rather than Gaussians.
Voigt profiles can be especially important when lines are saturated,
so their use is critical in modeling the wind observed in
GRO~J1655$-$40.  There are other minor differences between the codes,
including the elements and lines present in each code, but a full
review of each code is beyond the scope of this paper.

Each of the XSTAR and Cloudy models described in this work were
generated using a modified version of our best-fit continuum spectrum
as the input flux.  In order to prevent an arbitrarily high flux at
low energy, the power-law component was truncated below 1~keV.  This
is a reasonable step as the power-law component likely arises through
Comptonization.  We generated sets of models based partially on the
density constraints given above and the parameters in Paper I.  These
models all have a few properties in common: a constant density profile
($n \propto r^{0}$), a density of $n = 10^{14.0}~{\rm cm}^{-3}$, a
turbulent velocity of $300$~km/s, and a line-of-sight covering factor
of unity.  These models generally assumed the abundances used in Paper
I.  The XSTAR tables generated were two-dimensional grids in
ionization parameter and column density, making these variables the
parameters constrained directly by fitting.  Radii from XSTAR modeling
were derived using the relation $\xi = L_{X} / n r^{2}$.  For
complementarity and to get a direct measure of inner radius, the
Cloudy tables were generated as two-dimensional grids in radius and
column density.  These parameters were constrained by fitting, then,
and the ionization parameter was derived using the prior relation.

In order to assess whether or not magnetic processes drive the wind in
GRO~J1655$-$40, it is necessary to have a point of reference.  The
independent models developed in Netzer (2006) are consistent with
thermal driving.  For the pursposes of comparing our best--fit models
to a thermal wind prescription, then, we also created XSTAR and Cloudy
models based on the wind parameters detailed in Netzer (2006).
Special care was taken to match the thermal models in detail.  Whereas
XSTAR version ``ln'' was used to generate other models, we used a
modified XSTAR version ``kn'' that assumes an $r^{-2}$ density profile
to replicate the thermal models.  This version of XSTAR has a bug in
its treatment of Ni.  As a result, all models generated using XSTAR
version ``kn'' were set to have a Ni abundance of zero.  The radial
density profile of absorbing gas can be set directly in Cloudy, and so
the same version of Cloudy was used to generate all models.  All
models assumed a volume filling factor of $\Omega/4\pi = 0.2$ as per
Paper I, Netzer (2006), and the brief discussion above.  Netzer (2006)
actually describes two models with parameters that differ only
minimally.  We adopted a single set of parameters that accurately
represent the fundamentals of both models: ${\rm log}(\xi) = 3.0$,
${\rm log}(r_{0}) = 10.75$~cm, ${\rm log}(N) = 23.75$, $v_{turb} =
300$~km/s, ${\rm log}(n_{0}) = 13.0$, $n(r) \propto r^{-2}$, a line of
sight covering factor of 0.75, and solar abundances for all elements
apart from Ca and Fe which were given twice solar abundances.

All of the photoionization models were fit to the data using ISIS as
multiplicative table models modifying a disk blackbody plus power-law
continuum.  The neutral column density and power-law index were fixed
in such fits, but all other parameters were allowed to vary.  In one
sense, this methodology is artificially lax, since the photoionization
models were generated using the best-fit continuum model.  Presumably,
the best-fit continuum model could be enforced.  A more pragmatic
concern, though, is to assess how well the lines are fit rather than
how well a fiducial continuum can cope with calibration uncertainties
in specific wavelength bands.  For this purpose, allowing the
continuum parameters to vary is the only logical choice.  A wavelength
shift parameter is a part of all models.  It is clear from Table 1 and
Table 2 that there is a wide range in velocity shifts; therefore, this
parameter was also allowed to vary in all fits.  

In order to best evaluate the characteristics of the wind, we adopted
three families of fits aimed at removing any possible biases
associated with wavelength and/or charge state:\\

\noindent $\bullet$ One set of fits was made to the Fe~K band in the
1.3--2.5\AA~ range in order to focus on He-like Fe XXV and H-like Fe
XXVI.  These transitions survive even in extremely high ionization
regimes, and may therefore give the best information on the innermost
radial extent of the wind.

\noindent $\bullet$ A second set of fits were made to the
1.3--13.3\AA~ band.  The long wavelength limit was set by the cut-off
in the HEG spectra, and to avoid errors related to minor differences
between the HEG and MEG spectrometers.  This set of fits is meant to
assess the ``average'' properties of the wind, and to determine which
model best describes the spectrum as a whole.

\noindent $\bullet$ A final set of fits was made to narrow 2\AA~
slices between 1.3 and 13.3\AA~ in order to assess whether the wind
changes considerably with wavelength.  In essence, these fits test the
possibility that one (perhaps extreme) set of parameters is needed to
describe a small segment of the data, while other (perhaps less
extreme) parameters are sufficient to describe the bulk of the
observed absorption.\\

Table 3 summarizes the results of fits to the Fe K region.  Models 1A
and 2A are the fiducial thermal driving models based on the parameters
reported by Netzer (2006).  Models 1B and 2B assume an inner detection radius
equal to $0.01\times R_{C}$ ($10^{9.7}$~cm) -- 10 times smaller than
the radius at which thermal winds are possible.  These models are vast
statistical improvements over models 1A and 2A.  Models 1C and 2C are
the best-fit models for the Fe K band.  The inner radius was allowed
to float freely in these models; the fact that these models are
7$\sigma$ and 6$\sigma$ improvements over 1B and 2B, respectively,
likely signals that radii significantly less than even $0.01\times
R_{C}$ are statistically required.  Figures 5 and 6 compare the fiducial
thermal models to the best-fit models.

Clearly, even the ``best-fit'' models are not perfect descriptions of
the 1.3--2.5 \AA~ spectrum.  For instance, XSTAR does not include data
for lines from Mn, Cr, and Ti, and so these lines are not matched by
the models in Figure 5.  However, the overall fit is very good, and
the fact that models with a smaller inner wind radius and higher
density give superior fits is common to both XSTAR and Cloudy.  The
high ionization parameters measured ensure an Fe XXV/XXVI line ratio
that matches the data better than the fiducial thermal models.  In
those models, the low ionization parameter resulting from a low
density and large radius creates a large Fe XXV/XXVI line ratio that
fails to match the data.

Table 4 summarizes the results of fits to the broad 1.3--13.3 \AA~
band.  As with the fits made to the narrow Fe K band, the fiducial
thermal--driving models over-predict the strength of many lines, and
predict a number of lines that are not seen.  Models consistent with
an inner radius of $0.01\times R_{C}$ (models 3B and 4B) provide
vastly improved fits to the spectrum.  Models 3C and 4C are the
best-fit XSTAR and Cloudy models; they represent improvements over the
fiducial thermal--driving models at vastly more than the 8$\sigma$
level of confidence, again signaling that even smaller radii and
higher ionization parameters are statistically required.  The fiducial
thermal models and best--fit models are compared in Figures 8 and 9.

Examination of these figures shows that the best-fit models for the
1.3--13.3 \AA~ band fail to fit some lines.  Indeed, there are lines
that are not fit, or are fit poorly, by all of the models considered.
In particular, the complex 6--7 \AA~ range is not fit well; it is
possible that increasing the abundance of Mg may provide a slightly
better fit.  Despite these difficulties, models with a small inner
wind radius, high ionization parameter, and high density provide the
best overall description of the data.  Our fiducial thermal models
appear to not only fail to describe the Fe K band, but the broad-band
spectrum as well.  Netzer (2006) only reported on models for the
1.3--9.0\AA range; when models based on the same parameters are
extrapolated to 13.3\AA, many strong absorption lines are predicted
that could begin to obscure the continuum.  Again, all the above
findings are common to models generated with both XSTAR and Cloudy.

Finally, we made independent fits to six wavelength slices in the
1.3--13.3 \AA~ range to see if our fiducial thermal--driving model is
a better description of the data in any specific portion of the
spectrum.  Figures 10, 11, 12, and 13 show the results of fitting
these models to the data.  The panels in each of these figures include
the relevant parameters to facilitate direct slice-by-slice
comparisons.  The XSTAR implementation of the fiducial
thermal--driving models is shown in Figure 10, and the best-fit XSTAR
model is shown in Figure 11.  The Cloudy implementation of the
fiducial thermal--driving models are shown in Figure 12, and the
best-fit Cloudy models are depicted in Figure 13.  In each slice,
models with a small inner radius and high density provide vastly
better fits to the data.  This is true both in a statistical sense,
and from visual inspection.  These results are common to models
generated with both XSTAR and Cloudy.  The best fit model for each
slice gives an inner wind radius equal to or significantly within
$0.01\times R_{C}$.

The XSTAR model for the 11.3--13.3 \AA~ slice showing in Figure 11
provides the best overall fit to the density--sensitive Fe XXII lines
at 11.77 and 11.92\AA.  Note that this model also fits the putative Fe
XXII line at 11.47 \AA~ very well.  The model generated using Cloudy
appears to give two lines with approximately the right strength, but
the wavelengths look to be slightly in error.  (The nearby Ne IX 1s-2p
line is over-predicted, possibly signaling that an overabundance of
2.0 was too high.)  In contrast, even though the assumed partial
covering acts to reduce the depth of lines relative to the continuum,
the models based on Netzer (2006) predict lines that are far too
strong to match the data.  The XSTAR models for the Fe XXII lines are
shown in detail in Figure 14.  

In the exercises detailed above, the models consistent with thermal
driving sometimes appear to fail partially due to a poor continuum
fit.  The poor continuum fits are caused by the extremely strong lines
predicted by the self-consistent modeling procedure.  In fact, in all
cases the continuum parameters (apart from the power-law index) were
allowed to float in order to allow the best possible fit.  A procedure
that enforced a continuum more consistent with the best broad--band
continuum would cause models consistent with thermal driving to return
even worse fits to the data.

Folding together the results of the three fitting exercises: models
generated with XSTAR and Cloudy independently confirm that a high
density ($n \leq 10^{14.0}~{\rm cm}^{-3}$), highly ionized ($10^{4.5}
\leq \xi \leq 10^{5.4}$) wind detected close to the black hole
($10^9.0$--$10^{9.4}$~cm) provide a vastly superior description of the
disk wind spectrum than models suggesting that the wind originates at
a larger radius consistent with thermal driving.  The latter models
require a low ionization parameter and large radius, and therefore
predict line strengths and ratios that fail to match the data.

\section{Discussion}

The sections above detail our efforts to analyze the disk wind
absorption spectrum in GRO~J1655$-$40.  Both simple phenomenological
and self-consistent photoionization modeling was performed.  Fits were
made to specific wavelength regions, to the broadband spectrum, and in
short incremental wavelength slices.  Below, we begin by summarizing
the properties of the wind based on our fits, and then discuss the
nature of the wind.\\

\noindent {\it Density}: From simple Gaussian modeling of the
density--sensitive Fe XXII lines, and from Monte Carlo simulations
with CHIANTI, it is clear that the density of the wind is at least
$10^{13.6}$--$10^{13.8}~ {\rm cm}^{-3}$.  This range is significantly
higher than the value of $n = 10^{13.0}~ {\rm cm}^{-3}$ assumed in
Netzer (2006).  As discussed above, this higher range is effectively a
lower limit, and fits with models assuming a density of $10^{14}~{\rm
cm}^{-3}$ yield excellent fits.  It is clear from Figures 10--14 that
a high density and high ionization are needed to match the Fe XXII
lines.\\
 
\noindent {\it Radius}: All of the fits we obtained through direct
fitting of the self-consistent photoionization models -- regardless of
the wavelength region, and regardless of the code -- point to an inner
wind detection radius of $r_{0} \simeq 10^{9.0}$--$10^{9.4}$~cm.  BMS
and Woods et al.\ (1996) suggest that thermally driven winds are
possible for $r_{0} \geq 0.1$--$0.2~R_{C}$; in the case of
GRO~J1655$-$40, $0.1\times R_{C} \simeq 10^{10.7}$~cm.  Our fits
clearly show that an inner detection radius of $10^{9.7}$~cm, or
$0.01\times R_{C}$, provides a much better description of the data, in
both a statistical and visual sense.  Indeed, radii significantly less
than $10^{9.7}$~cm are statistically required.  An inner detection
radius of $10^{9.0}$--$10^{9.4}$~cm is 20--50 times smaller than the
radius at which BMS and Woods et al.\ (1996) suggest a thermal wind
can be driven.  These radii correspond to 500--1250~$R_{Schw.}$ for
GRO~J1655$-$40.  The wind may originate at a much smaller radius, but the
correspondingly higher ionization parameter would prevent detection of
the wind in that regime.\\

\noindent {\it Thickness}: The radial thickness of the wind can be
estimated via the relation $\Delta(r) = N_{H} / n$.  Models 1C, 2C,
3C, and 4C in Tables 3 and 4 then yield thicknesses of 1--5$\times
10^{9}$~cm.  Whereas we might trust Models 1C and 2C to give more
accurate estimates of the inner detection radius $r_{0}$, the
broadband fits made with models 3C and 4C may give a better measure of
thickness, so $5\times 10^{9}$~cm is likely the best value.  Thus, the
thickness of the wind is at least a few times the distance between the
central engine and the inner detection radus $r_{0}$.  The depth may
be larger if the wind originates within the radius at which detection
becomes possible.  If we were to instead insist that the wind fills
all of the interior radial distance with an $r^{-2}$ density law, then
$\xi = L_{X} / N_{H} r$ gives the radial thickness of the wind; values
of 1--6$\times 10^{9}$~cm are obtained.\\

\noindent {\it Mass Flux}:  The mass flux predicted by the thermal
wind models is far lower than observed, even if the gas were
beyond 0.1 $R_{C}$ (see below).  Equation 4.8 of Woods et
al. (1996), or their Figure 16 scaled from a $10^{8}~M_{\odot}$ black
hole to a 7 $M_{\odot}$ black hole, predict the peak mass loss rate
per unit area.  For an ionizing spectrum that gives a Compton
temperature close to that found in our models or Netzer's, the maximum
is $6 \times 10^{-6} \rm g~cm^{-2}~s^{-1}$, compared to and observed
mass flux of about $10^{-2} \rm g~cm^{-2}~s^{-1}$.  In terms of
densities, the peak mass flux predicted by the model divided by the
500 $\rm km~s^{-1}$ flow speed along the line of sight gives a density
of $6 \times 10^{10}~\rm cm^{-3}$.  Woods et al. find that the flow is
nearly vertical between 0.1 $R_{C}$ and $R_{C}$, and the high
inclination angle means that the actual speed is considerably larger
than the line-of-sight component, so the predicted density is still
smaller.  Therefore, the thermally driven models predict densities and
mass fluxes three orders of magnitude smaller than required by the
observed wind spectrum.  

In terms of total mass outflow rate, the best-fit values for density,
inner detection radius, and thickness give a value of 3--6$\times
10^{17}$~g/s.  Here again, this estimate should be regarded as a lower
limit, because the velocity into our line of sight may be a small
fraction of the velocity perpendicular to the disk and the wind may
originate at a smaller radius.  Assuming that accretion onto
GRO~J1655$-$40 has an efficiency of 10\%, the mass accretion rate
through the disk is approximately $5.6\times 10^{18}$~g/s.  The mass
flux in the wind is at least 5--10\% of that being accreted, and could
easily be comparable to the mass accretion rate through the disk.

\noindent{\it Kinetic Energy Flux}: The numbers above give a kinetic
energy flux of 0.5--1.0$\times 10^{13}~ {\rm erg}~ {\rm cm}^{-2}~ {\rm
s}^{-1}$.  An angle of $9^{\circ}$ above the disk midplane (see above)
translates into a height of $Z = 0.16r$.  For an average wind
detection radius of $r_{avg} = 2\times 10^{9}$~cm, the energy flux
required to lift the gas to that height is 3--6$\times 10^{14}~ {\rm
erg}~ {\rm cm}^{-2}~ {\rm s}^{-1}$.  As the wind can originate at a
radius smaller than the detection radius, the kinetic energy flux
could also be larger than the simple estimate given here.  For
comparison, the viscous energy flux dissipated is given by $3 G M_{BH}
\dot{m}_{acc} / 8\pi r^{3} = 8.0\times 10^{16} {\rm erg}~ {\rm
cm}^{-2}~ {\rm s}^{-1}$.\\

In the analysis above, three independent photoionization codes and a
number of unbiased fitting experiments agree: the wind observed in
GRO~J1655$-$40 originates at least an order of magnitude (and probably
more than 20--50 times closer) to the black hole than is predicted by
thermal driving models, and is three orders of magnitude denser than
predicted by thermal driving models.  It is possible that the seminal
works on Compton-heated winds in accreting systems can be revised, but
it is unlikely that the full extent of these disparities can be easily
overcome.  Given that radiative driving also fails, the only realistic
possibility remaining is that the disk wind in GRO~J1655$-$40 is
driven in part by magnetic processes.

A brief exploration of the reasons why models consistent with thermal
driving might have previously appeared favorable is now in order.  The
analysis presented by Netzer (2006) broadly agreed with many of the
conclusions of Paper I regarding elemental abundances, the inadequacy
of radiation pressure driving, and the limited solid angle of the
absorbing gas as seen from the black hole.  But whereas Netzer (2006)
minimized $\chi^{2}$ for a large number of absorption lines, Paper I
demanded that the model match the lowest and highest charge states of
Fe, placing a more stringent limit on the ionization parameter.  The
latter procedure may be better for determining the innermost extent of
the wind.  More importantly, Netzer (2006) argued for saturation of
one member of the density-sensitive line pair of Fe XXII lines,
permitting a smaller density.  As shown above, the Fe XXII lines are
not saturated, and a higher density is required.  Assuming a
line-of-sight covering factor less than unity diminished the depth of
lines relative to the continuum models, perhaps slightly improving the
way that a combination of low density, low ionization, and large
radius fit the data.  In fact, the data likely rule out partial
covering.  Finally, the assumption of Gaussian line profiles may have
affected Netzer's modeling.  When saturation is important (as with the
Fe XXV and XXVI lines, among others), Voigt profiles are required.
The photoionization model in Paper I, and the all of the models used
in this work, used Voigt profiles.

Blandford \& Payne (1982) have noted that if magnetic field lines are
anchored in a disk, centrifugal forces will allow gas to escape along
field lines.  This magnetocentrifugal wind is the kind found in FU Ori
stars, so there is an observational confirmation of this mechanism.
Tangential velocities are not encoded into X-ray absorption lines,
so it is difficult to reveal clear signatures of
rotation.  However, geometric considerations may make it possible to
discern whether this mechanism, or another magnetic mechanism, is
driving a wind.  The optimal configuration for powering winds in this
way is one where field lines rise off the plane of the disk at an
angle of about 60$^{\circ}$ (Spruit 1996).  If ionized X-ray winds in
X-ray binaries are primarily magnetocentrifugal, strong wind
signatures should be seen for a wide range of binary inclinations.

Another possibility is that the Poynting flux generated from the
magnetic viscosity thought to drive accretion through a disk can power
a wind.  Recent work has shown that the magneto-rotational-instability
(MRI) can drive turbulence, viscosity, and accretion through disks
(see, e.g., Balbus \& Hawley 1991; Hawley, Gammie, \& Balbus 1995;
Balbus \& Hawley 1998; Krolik, Hawley, \& Hirose 2005).  Indeed, the
original $\alpha$-disk prescription written down by Shakura \& Sunyaev
(1973) envisaged a source of internal magnetic viscosity like MRI.
Disk simulations incorporating MRI show that this mechanism can
transmit 25\% of the magnetic energy flux out of the disk to power a
wind (Miller \& Stone 2000), though Blaes (2007) finds much smaller
values.  In the case of GRO~J1655$-$40, even the more conservative
estimate of the magnetic energy flux is more than adequate to drive the
observed wind to infinity, even if the true outflow rate is much
larger because the flow is more nearly perpendicular to the face of
the disk.

It is important to note that at the radii suggested by photoionization
modeling, the blue-shifts observed in GRO~J1655$-$40 are about an
order of magnitude below the local escape velocity.  X-ray absorption
in black holes is only sensitive to velocities along our line of sight
to the central engine.  The wind may retain some of the circular
motion of the disk, but this cannot be detected in absorption.
Similarly, there is also good reason to think that the wind may be
launched vertically from the disk; the velocity we measure along our
line of sight would then be a small fraction of the total.  As noted
above, simulations suggest that MRI viscosity can supply the magnetic
energy flux needed to drive the wind to infinity in this circumstance.
The lack of emission lines from material at greater heights above the
disk would naturally be explained by acceleration causing a drop in
density and an increase in the ionization parameter.  Thus, the data
are consistent with the wind escaping to infinity, and this appears to
be the most likely scenario.  Even if the wind does not escape the
system, but is instead a local flow stemming from a turbulent magnetic
region above the disk, the implications for disk accretion onto black
holes are just as interesting.

Perhaps the most detailed picture of winds driven by pressure from
internal magnetic viscosity is given by Proga (2003).  This work
predicts that magnetic winds are likely to be dense along the plane of
the disk, and to have low outflow velocities.  The observed
properties of the wind in GRO~J1655$-$40 match these general
predictions very well.  Whereas the field lines in magnetocentrifugal
winds are largely poloidal, the field lines are predominantly toroidal
in this case, and the gradient in the toroidal field supplies
pressure.

It is not possible to clearly distinguish which magnetic mechanism is
driving the wind in GRO~J1655$-$40, but the data and current models
would seem to favor MRI pressure.  The wind may not lie exclusively
along the plane of the disk, but its ionization parameter will
increase with any fall-off in density well above the disk, making it
difficult or impossible to detect.  The dense, observable part of the
wind, at least, is constrained to lie along the plane of the disk.
The low outflow velocity is also in keeping with predictions.  Proga
(2003) finds that the rigid, largely poloidal field structure required
for magnetocentrifugal driving is very difficult to maintain in dense
winds.  Recent work on magnetocentrifugal winds tailored to T Tauri
stars, however, finds that such winds may persist even when mass
loading is high and the toroidal component of the field dominates
(Anderson et al.\ 2005).

Some additional insights into this question might be gleaned by briefly
examining winds in other stellar-mass black holes.  GX~339$-$4 is
likely viewed more closely to face-on than GRO~J1655$-$40, and the
wind observed in that source is less dense than in GRO~J1655$-$40, it
has a much lower ionization parameter, and lower mass outflow rate
(Miller et al.\ 2004).  GRS~1915$+$105 is seen at a high inclination,
like GRO~J1655$-$40, and a highly ionized wind is seen in this source
(Lee et al.\ 2002).  The black hole candidates H1743$-$322 and 4U
1630$-$472 do not have inclinations constrained through optical or IR
observations, but dipping behavior in these sources demands a high
inclination in both cases.  In H 1743$-$322, a highly ionized wind has
also been observed; photoionization modeling suggests an inner
launching radius of $r_{0} = 1\times 10^{8}$~cm (Miller et al.\ 2006).
Recent observations of 4U~1630$-$472 have also revealed a highly
ionized wind originating close to the black hole (Kubota et al.\
2007).  The winds in these sources may also be driven by magnetic
processes.  

Disk winds in stellar-mass black holes appear to be stronger in soft
disk--dominated states than in other states.  GRO~J1655$-$40 was
observed with the {\it Chandra}/HETGS on a separate occasion during
the 2005 outburst, but in a spectrally harder state.  Only the Fe XXVI
line is clearly detected in that observation (a full analysis will be
presented in a separate paper).  H 1743$-$322 was observed on four
occasions with the {\it Chandra}/HETGS, and the only observation
wherein Fe XXV and Fe XXVI absorption lines are {\it not} detected is
in the one observation made in a spectrally hard state.  A preliminary
analysis of archival {\it Chandra}/HETGS observations of
GRS~1915$+$105 again suggests that lines are strongest in soft states,
and diminished in hard states.  The higher ionizing flux produced
by the central engine in hard states acts to drive-up the ionization
parameter and reduce the number of lines, but geometric changes and/or
changes in the ability of the disk to drive a wind (perhaps due to
changes in the magnetic field configuration) appear to be important
too (Miller et al.\ 2006).  

If winds are truly stronger in soft states, a number of interesting
possibilities and consequences arise.  Jets are quenched in soft
disk--dominated states (e.g., Fender 2006), but present in low--flux,
spectrally hard states.  It is possible that winds and jets are
anti-correlated, and that the disk alternates its outflow mode.  For
instance, if winds are typically very dense (as in soft states),
perhaps a magnetic field configuration conducive to jets (a sort of
extreme Blandford \& Payne scenario) is impossible.  But if the wind
density drops, perhaps the jet--producing configuration becomes possible.

Though GRO~J1655$-$40 may be the best example apart from FU Ori (and
perhaps T Tauri) stars, there is already evidence that winds may be
driven by magnetic processes in some white dwarfs and AGN (Mauche \&
Raymond 2000, Kraemer et al.\ 2005).  In black holes and neutron
stars, one hallmark of such winds may be the presence of Fe XXV and Fe
XXVI absorption lines.  It appears difficult to produce Fe XXV and
XXVI lines of comparable strength in gas with an ionization parameter
below $10^{3}$; a value of $\xi \simeq 10^{4}$ seems more common.
This typically implies a region that is very close to the central
engine.  The principal difficulty in searching for these lines is
sensitivity: source spectra and instrument effective area curves both
fall rapidly through the Fe~K band.  Whereas few stellar-mass black
holes are persistent, many neutron star binaries are peristent sources
and therefore attractive targets.  Unfortunately, neutron stars are
fainter than black holes at the same Eddington fraction, and their
spectra are softer.  Thus, whether looking for ionized winds in
stellar-mass sources or AGN, deep spectroscopic observations are
required.  For sufficiently bright sources, the {\it Chandra}/HETGS is
the best instrument for this purpose, but for dimmer sources the
combination of high collecting area and favorable CCD modes aboard
{\it Suzaku} may offer a pragmatic way to search for ionized winds.

Time-resolved high-resolution spectroscopy of CVs has allowed
observers to identify different origins for the nature of various
absorption lines, and to test how winds are driven in these systems
(see, e.g., Hartley, Drew, \& Long 2002; Hartley et al.\ 2002).
Longer observations which sample large fractions of a binary phase in
multiple outburst states may provide an improved means of testing the
nature of winds in stellar-mass black holes and neutron stars.
Whereas the properties of thermal winds should depend strongly on the
launching radius and illuminating flux, the velocities and column
densities of lines originating in magnetic winds may relatively
independent of modest changes in the central engine.  Especially since
many sources are transient, such observations will be difficult to
schedule and execute, but the dividends for our understanding of disk
accretion onto black holes and neutron stars may be well worth the
effort.

\section{Conclusions}

We have analyzed a {\it Chandra}/HETGS spectrum of the stellar-mass
black hole GRO~J1655$-$40.  A highly ionized accretion disk wind is
revealed in great detail.  Both simple considerations and advanced
modeling with three independent photoionization codes suggest that
radiative and thermal driving mechanisms are likely ruled out.  The
wind originates at least an order of magnitude too close to the black
hole (and likely 20--50 times too close), and is three orders of
magnitude too dense, for it to be a thermal wind of the kind described
in seminal treatments.  The wind is more likely powered by magnetic
driving.  New models for magnetic disk winds make predictions that
match the data, at least qualitatively.  At present, theoretical
studies provide the strongest support for MRI operating in black hole
disks.  Observational evidence that disk accretion onto compact
objects is facilitated by magnetic processes has been elusive; this
observation may provide indirect observational support.

\vspace{0.25in}

We wish to thank Harvey Tananbaum, Jean Swank, and the {\it Chandra}
and RXTE teams for executing our TOO observations.  JMM is grateful to
Gary Ferland and Ryan Porter for their patient help and insights, and
hospitality during a productive visit to the University of Kentucky.
We thank Mitch Begelman, Roger Blandford, Ed Cackett, Nuria Calvet,
Lee Hartmann, Michiel van der Klis, Jerry Orosz, Daniel Proga, Michael
Rupen, Mateusz Ruszkowski, Danny Steeghs, Marta Volonteri, and Rudy
Wijnands for helpful discussions.  We thank the anonymous referee for
helpful comments which improved the paper.  JMM acknowledges funding
awarded to support this research through the {\it Chandra} guest
observer program.  This work made use of the facilities and tools
available through the HEASARC web page, operated by GSFC for NASA.

\pagebreak

\centerline{~\psfig{file=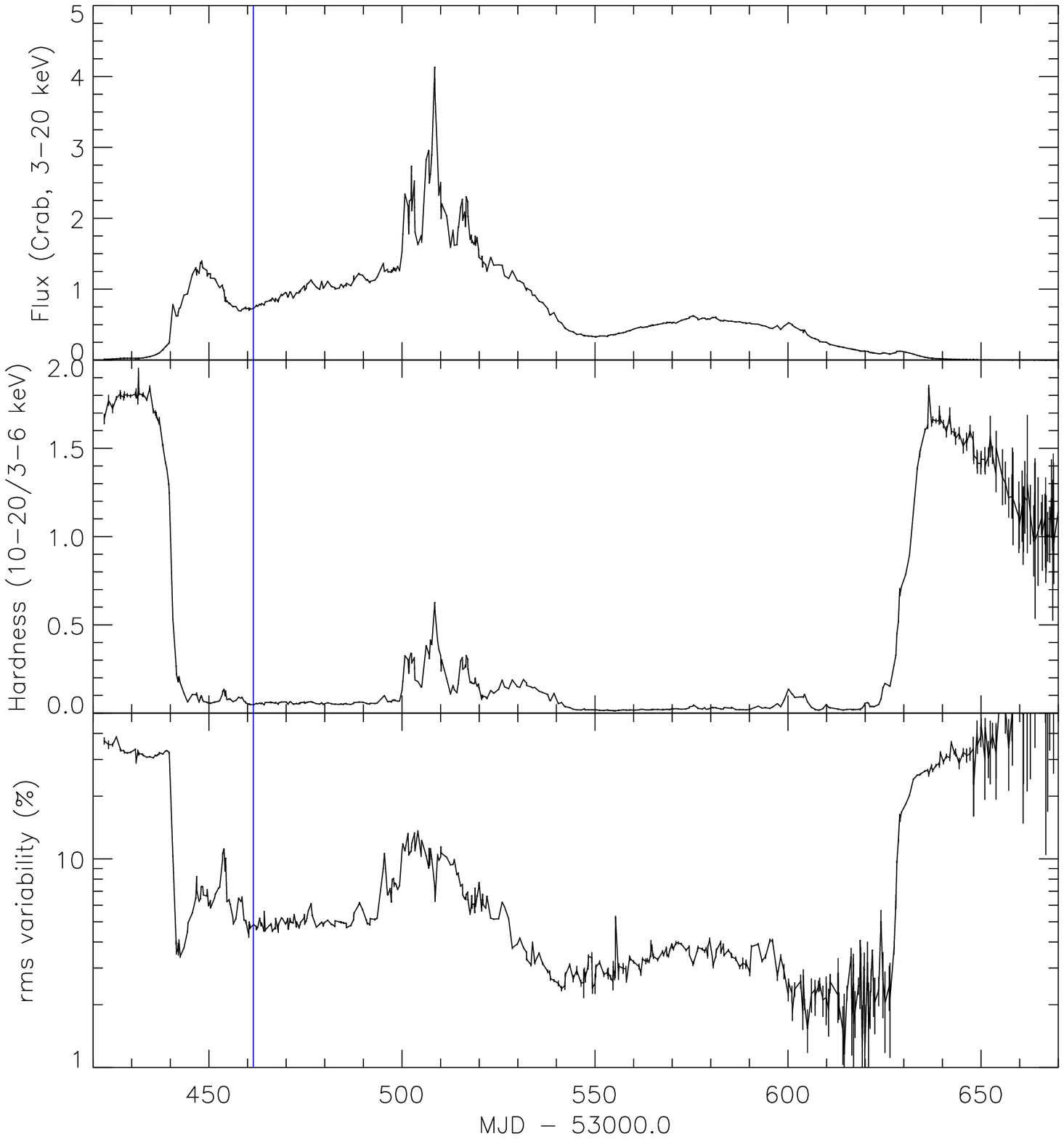,width=4.0in}~}
\figcaption[h]{\footnotesize {\it RXTE} made numerous pointed
observations of GRO J1655$-$40 during its 2005 outburst.  Flux (in
Crab units), spectral hardness, and rms variability are plotted versus
time in the figure above, based on these pointed observations.  The
 blue vertical line in each panel shows the time at which the
{\it Chandra} observation was obtained.  (Error
bars are plotted, but are often very small.)}
\medskip

\centerline{~\psfig{file=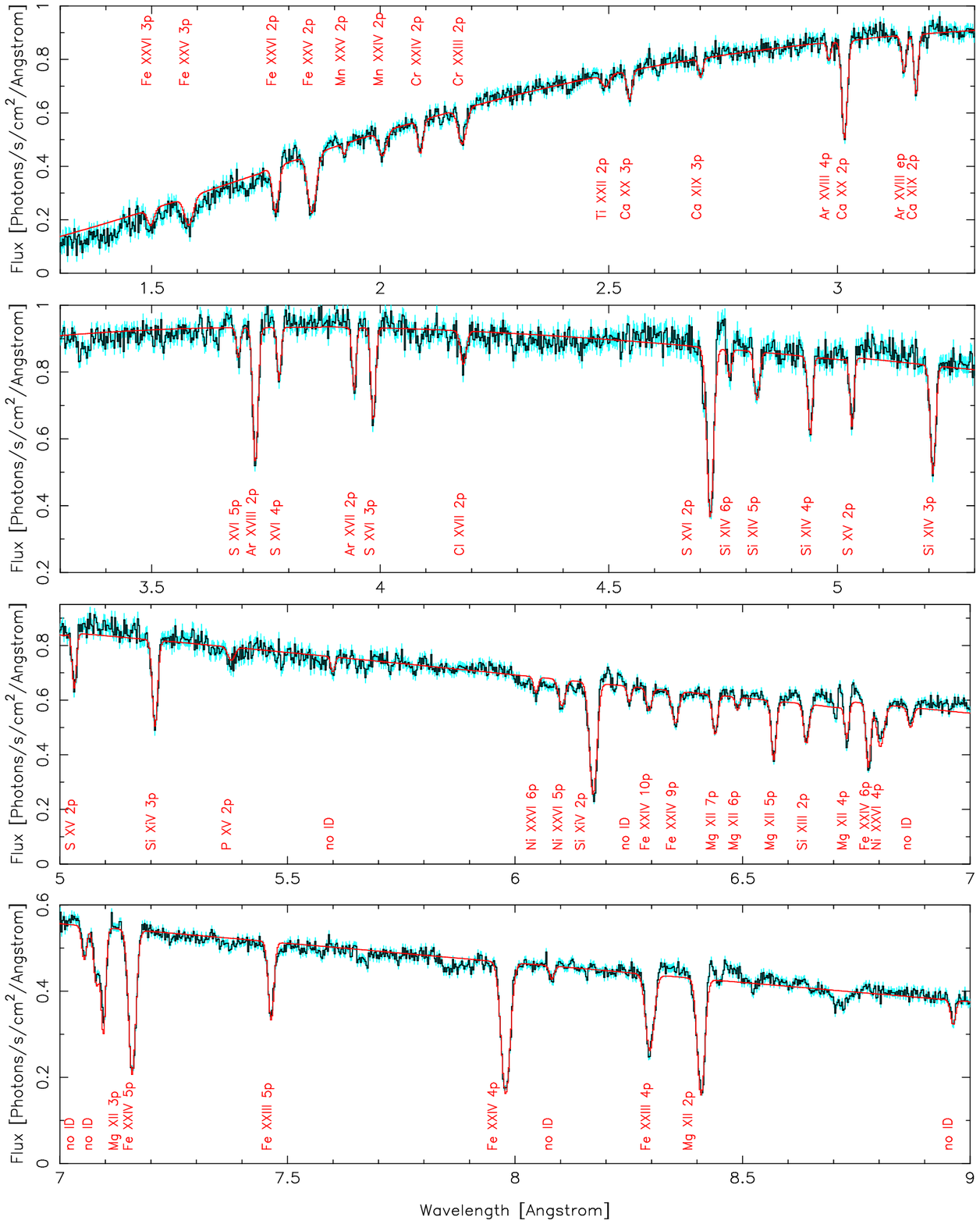,width=6.5in}~}
\figcaption[h]{\footnotesize Phenomenological fits to the {\it
Chandra}/HETGS spectrum of GRO~J1655$-$40 are shown above.  The
continuum was only fit to the wavelenth slice shown in each panel, and
all lines were fit with simple Gaussian functions.  Line
identifications are given based on the strategy outlined in the text.}

\centerline{~\psfig{file=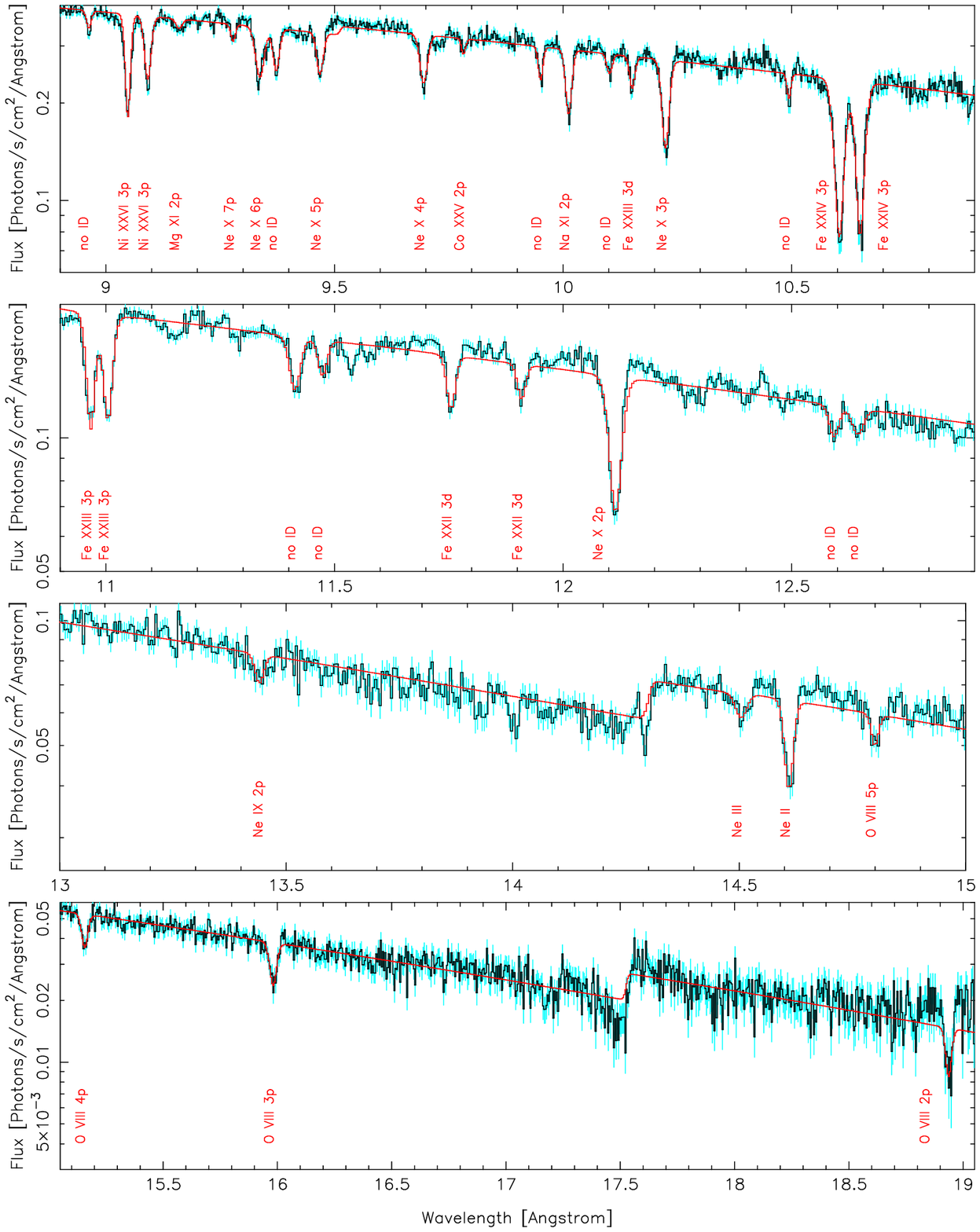,width=6.5in}~}
\figcaption[h]{\footnotesize Phenomenological fits to the {\it
Chandra}/HETGS spectrum of GRO~J1655$-$40 are shown above.  The
continuum was only fit to the wavelenth slice shown in each panel, and
all lines were fit with simple Gaussian functions.  Line
identifications are given based on the strategy outlined in the text. }

\begin{table}[t]
\caption{Gaussian Line Parameters and Line Flux Limits, $Z<17$}
\begin{scriptsize}
\begin{center}
\begin{tabular}{lllllllllll}
\multicolumn{3}{l}{Ion and} & Meas. & Theor. & Shift & \multicolumn{2}{c}{FWHM} & Flux ($10^{-3}$ & W & N$_{\rm Z}~(10^{17}$ \\
\multicolumn{3}{l}{Transition} & (\AA) & (\AA) & (km/s) & ($10^{-3}$\AA) & (km/s) & ph/cm$^{2}$/s) & (m\AA) & ${\rm cm}^{-2}$) \\
\hline

\multicolumn{3}{l}{O VIII $1s-5p$} & 14.800(2) & 14.8206 & 420(40) & $<10$ & $<200$ & 1.4(3) & 23(5) & 9(2) \\

\multicolumn{3}{l}{O VIII $1s-4p$} & 15.157(2) & 15.1762 & 380(40) & 28(5) & 550(80) & 4.0(4)	& 77(8)	& 13(1) \\

\multicolumn{3}{l}{O VIII $1s-3p$} & 15.982(2) & 15.987 & 90(30) & 31(5) & 580(90) & 	5.1(5) & 120(10)  & 7.5(8) \\

\multicolumn{3}{l}{O VIII $1s-2p$} & 18.938(2) & 18.9689 & 480(40) & 24(5) & 380(80) & 1.2(2)	& 80(8)	& 0.7(1) \\

\multicolumn{3}{l}{Ne X $1s-7p$} & 9.2786(6) & 9.2912	& 410(20) & 5(2) & 200(100) & 1.3(2) &	3.6(7) &  10(2) \\

\multicolumn{3}{l}{Ne X $1s-6p$} & 9.3365(6) & 9.3616	& 800(20) & 22(1) & 700(40) & 6.6(3) & 22(1) & 35(1) \\

\multicolumn{3}{l}{Ne X $1s-5p$} & 9.4685(4) & 9.4807	& 390(20) & 14(1) & 440(30) & 4.1(3) & 12(1) & 11(1) \\

\multicolumn{3}{l}{Ne X $1s-4p$} & 9.6949(5) & 9.7082	& 410(10) & 18(2) & 560(50) & 5.5(3) & 16(1) & 6.6(3) \\

\multicolumn{3}{l}{Ne X $1s-3p$} & 10.2246(3) & 10.2389 & 420(10) & 17(1) & 500(30) &	8.3(2) & 30(7) & 4.1(1) \\

\multicolumn{3}{l}{Ne X $1s-2p$} & 12.1152(5) & 12.1330 & 550(10) & 29(1) & 720(20) & 8.2(4) & 57(3) & 1.05(5) \\

\multicolumn{3}{l}{Ne IX $1s^{2}-1s2p$} & 13.441(4) & 13.4471	& 140(80) & 14 & 300 & 1.8(4) &	22(5) &	0.20(5) \\

\multicolumn{3}{l}{Ne III} & 14.504(3) & 14.526 & 450(50) & 20(10) & 400(200) & 1.5(2) & 22(3) & 1.1(2) \\

\multicolumn{3}{l}{Ne II} & 14.611(1) & 14.631 & 410(20) & 14(2) & 300(30) & 3.8(3) & 58(5) & 4.9(4) \\

\multicolumn{3}{l}{Na XI $1s-2p$} & 10.0122(3) & 10.0250 & 380(10) & 13.2(5) & 400(20) & 5.6(2)	& 18.7(7) & 0.51(2) \\

\multicolumn{3}{l}{Mg XII $1s-7p$} & 6.4395(3) & 6.4486 & 420(10) & 9(1) & 420(50) & 3.2(1) &	5.1(2) & 28(1) \\

\multicolumn{3}{l}{Mg XII $1s-6p$} & 6.4888(5) & 6.4974	& 400(20) &  $<6$ & $<300$ & 0.8(1) & 1.3(2) &	4.4(6) \\

\multicolumn{3}{l}{Mg XII $1s-5p$} & 6.5685(2) & 6.5801	& 530(10) &  8.0(5) & 360(30) &	4.7(1) & 7.7(2) & 14.4(4) \\

\multicolumn{3}{l}{Mg XII $1s-4p$} & 6.7275(3) & 6.7379	& 460(10) &  $<6$ & $<300$ & 2.5(1) & 4.2(2) & 3.6(2) \\

\multicolumn{3}{l}{Mg XII $1s-3p$} & 7.0969(4) & 7.1062	& 390(20) &  5.4(5) & 230(20) & 3.9(3) & 9.8(9)	& 2.8(3) \\

\multicolumn{3}{l}{Mg XII $1s-2p$} & 8.4087(2) & 8.4210	& 440(10) & 16.7(3) & 590(10) & 9.6(1) & 22.6(2) & 0.87(1) \\	

\multicolumn{3}{l}{Mg XI $1s^{2}-1s2p$}	& 9.159(2) & 9.1688 & 320(70) & 13 & 400 & 1.0(1) & 2.7(3) & 0.049(5) \\

\multicolumn{3}{l}{Si XIV $1s-6p$} & 4.764(2) & 4.7704 & 400(100) & $<5.0$ & $<300$ & 1.0(1) & 1.2(1) & 7.5(8) \\

\multicolumn{3}{l}{Si XIV $1s-5p$} & 4.8243(7) & 4.8311	& 420(40) & 9(2) & 600(100) & 2.4(2) & 2.8(2) & 9.7(9) \\	

\multicolumn{3}{l}{Si XIV $1s-4p$} & 4.9407(3) & 4.9469	& 380(20) & 6(1) & 360(60) & 3.6(2) & 4.2(2) & 6.7(3) \\

\multicolumn{3}{l}{Si XIV $1s-3p$} & 5.2090(3) & 5.2172	& 470(20) & 9.7(7) & 560(40) & 5.9(2) &	7.2(2) & 3.8(1) \\

\multicolumn{3}{l}{Si XIV $1s-2p$} & 6.1721(1) & 6.1822	& 490(10) & 15.5(2) & 750(10) &	11.9(1)	& 18.5(2) & 1.32(1) \\

\multicolumn{3}{l}{Si XIII $1s^{2}-1s2p$} & 6.6402(2) & 6.6480 & 350(10) & 10(1) & 450(50) & 3.4(1) & 5.7(1) & 0.19(1) \\

\multicolumn{3}{l}{P XV $1s-2p$} & 5.375(2) & 5.383 & 700(200) &  -- & 400 & 1.0(2) & 1.3(3) & 0.12(3) \\

\multicolumn{3}{l}{S XVI $1s-5p$} & 3.690(1) & 3.6959 &  480(80) & $<5$	& $<400$ & 1.1(2) & 1.2(2) & 7(1) \\

\multicolumn{3}{l}{S XVI $1s-4p$} & 3.780(1) & 3.7845 & 360(80)	&  $<5$	& $<400$ & 1.9(2) & 2.0(2) & 5.4(5) \\

\multicolumn{3}{l}{S XVI $1s-3p$} & 3.9858(3) & 3.9912 & 400(30) & 6(1)	& 450(80) & 3.9(2) & 4.2(2) & 3.8(2) \\
  
\multicolumn{3}{l}{S XVI $1s-2p$} & 4.7221(2) & 4.7292 & 450(10) & 14.4(7) & 910(50) &	11.3(2) & 12.8(2) & 1.55(2) \\

\multicolumn{3}{l}{S XV $1s^{2}-1s2p$} & 5.0318(4) & 5.0387 & 410(20) & $<5$ & $<400$ & 2.9(2) & 3.4(2)	& 0.20(2) \\

\hline
\end{tabular}
\vspace*{\baselineskip}~\\ \end{center} 
{Parameters of the disk wind absorption lines observed in
GRO~J1655$-$40 for elements with $Z<17$ are given above.  For clarity,
the lines are listed by element in order of ascending atomic number,
and in order of increasing wavelength by element and ion.  The
spectral continua were fit locally using power-law models modified by
neutral photoelectric absorption edges (due to the interstellar
medium) where appropriate.  The lines were fit with simple Gaussian
models.  The errors quoted above are $1\sigma$ uncertainties.  Line
significances were calculated by dividing line flux by its 1$\sigma$
error.  Where errors are not given, the parameter was fixed at the
quoted value.  Line widths consisent with zero are not resolved.}
\vspace{-1.0\baselineskip}
\end{scriptsize}
\end{table}

\pagebreak

\begin{table}[t]
\caption{Gaussian Line Parameters for $Z>17$ Elements and Unidentified Lines}
\begin{scriptsize}
\begin{center}
\begin{tabular}{lllllllllll}
\multicolumn{3}{l}{Ion and} & Meas. & Theor. & Shift & \multicolumn{2}{c}{FWHM} & Flux ($10^{-3}$ & W & N$_{\rm Z}~(10^{17}$ \\
\multicolumn{3}{l}{Transition} & (\AA) & (\AA) & (km/s) & ($10^{-3}$\AA) & (km/s) & ph/cm$^{2}$/s) & (m\AA) & ${\rm cm}^{-2}$) \\
\hline

\multicolumn{3}{l}{Cl XVII $1s-2p$} & 4.182(1) &  4.187	& 400(100) & 9(2) & 600(200) & 1.5(2) & 1.6(2) & 0.25(3) \\

\multicolumn{3}{l}{Ar XVIII $1s-4p$} & 2.981(1) & 2.9875 & 700(100) & $<5$ & $<500$ & 0.8(1) & 	0.9(1) & 3.9(5) \\

\multicolumn{3}{l}{Ar XVIII $1s-3p$} & 3.1454(5) & 3.1506 & 500(50) & 6(1) & 600(100) & 2.0(2) & 2.2(2)	& 3.2(3) \\

\multicolumn{3}{l}{Ar XVIII $1s-2p$} & 3.7271(2) & 3.7329 & 470(20) & 8.7(7) & 700(50) & 6.3(2)	& 8.0(3) & 1.6(1) \\

\multicolumn{3}{l}{Ar XVII $1s^{2}-1s2p$} & 3.9429(4) & 3.9488 & 450(30) & $<5$	& $<500$ & 2.3(1) & 2.6(1) & 0.24(1) \\

\multicolumn{3}{l}{K XIX $1s-2p$} & --	&   3.348 & -- &   -- & 800 & 1.0(2) & 1.0(2) & 0.24(5) \\

\multicolumn{3}{l}{Ca XX $1s-3p$} & 2.5452(6) & 2.5494	& 490(70) &  9(2) & 1100(100) & 1.7(2) & 2.2(2)	 & 4.8(5) \\

\multicolumn{3}{l}{Ca XX $1s-2p$} & 3.0187(2) & 3.0203	& 160(20) &  9.2(7) & 910(70) &	5.75(7)	& 6.66(8) &  2.0(1) \\

\multicolumn{3}{l}{Ca XIX $1s^{2}-1s3p$} & 2.701(1) & 2.7054 & 500(100) & $<10$	& $<1100$ & 0.9(1) & 1.1(1) & 1.1(1) \\

\multicolumn{3}{l}{Ca XIX $1s^{2}-1s2p$} & 3.1722(3) & 3.1772 &	470(30)	&  $<10$ & $<1100$ & 2.9(2) & 3.2(2) &  0.46(4) \\

\multicolumn{3}{l}{Sc XXI $1s-2p$} & --	&  2.740 & -- & -- & 1500 & $<0.1$ & $<0.13$ & $<0.05$ \\

\multicolumn{3}{l}{Ti XXII $1s-2p$} & 2.493(2) & 2.4966 & 430(240) & 17(5) & 2000(600) & 1.0(2) & 1.3(3) & 1.7(3) \\

\multicolumn{3}{l}{V XXIII $1s-2p$} &  -- & 2.2794 & --	& -- & 1500 & $<0.7$ & 	$<0.7$ &  $<0.4$ \\

\multicolumn{3}{l}{Cr XXIV $1s-2p$} & 2.0880(6) &  2.0901 & 300(80) & 10(2) & 1400(300)	& 2.0(2) & 3.4(3) & 6.3(6) \\

\multicolumn{3}{l}{Cr XXIII $1s^{2}-1s2p$} & 2.1794(6) & 2.1821	& 370(80) & 19(2) & 2600(300) & 3.1(2) & 4.8(3)	& 1.6(2) \\

\multicolumn{3}{l}{Mn XXV $1s-2p$} & 1.922(2) & 1.9247	& 400(300) & $<154$ & $<2000$ & 0.5(1) & 1.0(2) & 1.1(2) \\

\multicolumn{3}{l}{Mn XXIV$1s^{2}-1s2p$} & 2.005(1) &  2.0062 &	200(140) & 17(2) & 2500(300) & 2.0(2) & 3.7(4) & 3.7(4) \\

\multicolumn{3}{l}{Fe XXVI $1s-3p$} & 1.498(2) &  1.5028 & 1000(400)  & 12 & 2400 & 1.1(2) & 5(1) & 32(6) \\

\multicolumn{3}{l}{Fe XXVI $1s-2p$} & 1.7714(5) & 1.7798 & 1400(100) &  12(1) & 2400(200) & 3.1(2) & 7.8(5) & 6.7(4) \\

\multicolumn{3}{l}{Fe XXV $1s^{2}-1s3p$} & 1.581(1) & 1.5732 & 1500(200)  & 20 & 3800 & 2.6(2) & 9.7(9) & 28(3) \\

\multicolumn{3}{l}{Fe XXV $1s^{2}-1s2p$} & 1.8510(4) & 1.8504 &	0(100) &  20(1)	& 3800(300) & 5.4(2) & 12.6(5) & 5.2(2) \\

\multicolumn{3}{l}{Fe XXIV $1s^{2}2s-1s^{2}10p$} & 6.2946(5) & 6.3055 &	-- & 10(1) & 480(40) & 1.8(2) & 2.8(3) & 21(1) \\

\multicolumn{3}{l}{Fe XXIV $1s^{2}2s-1s^{2}9p$}	& 6.3523(4) & 6.3475 & -- & 14(1) & 660(50) & 3.3(2) & 5.2(3) & 28(2) \\

\multicolumn{3}{l}{Fe XXIV $1s^{2}2s-1s^{2}8p$} & blend & -- & -- & -- & -- & -- & -- & -- \\

\multicolumn{3}{l}{Fe XXIV $1s^{2}2s-1s^{2}7p$} & blend & -- & -- & -- & -- & -- & -- & -- \\

\multicolumn{3}{l}{Fe XXIV $1s^{2}2s-1s^{2}6p$} & 6.7773(2) & 6.7870 & 430(10) & 8.2(5) & 360(20) & 5.4(1) & 10.2(2) & 11.9(2) \\

\multicolumn{3}{l}{Fe XXIV $1s^{2}2s-1s^{2}5p$}	& 7.1590(1) & 7.1690 & 420(10) & 15.8(2) & 660(10) & 10.37(3) & 19.3(1)	& 10.6(1) \\

\multicolumn{3}{l}{Fe XXIV $1s^{2}2s-1s^{2}4p$} & 7.9795(1) & 7.9893 & 370(10) & 20.2(1) & 760(10) & 13.1(2) & 28.2(4) & 5.15(7) \\

\multicolumn{3}{l}{Fe XXIV $1s^{2}2s-1s^{2}3p$} & 10.6043(3) & 10.619 & 420(10)	& 27.3(7) & 770(20) & 13.7(4) & 60(1) & 2.30(4) \\

\multicolumn{3}{l}{Fe XXIV $1s^{2}2s-1s^{2}3p$} & 10.6494(3) & 10.663 & 380(10)	& 23.8(5) & 670(20) & 11.9(2) & 52(1) & 3.95(5) \\

\multicolumn{3}{l}{Fe XXIII $2s^{2}-2s5p$} & 7.4639(2) & 7.4722	& 330(10) & 11.6(5) & 470(20) & 4.15(4)	& 8.2(1) & 2.27(4) \\

\multicolumn{3}{l}{Fe XXIII $2s^{2}-2s4p$} & 8.2963(2) & 8.3029	& 240(10) & 18.1(5) & 650(20) &	7.0(1) & 16.3(2) & 1.49(2) \\

\multicolumn{3}{l}{Fe XXIII $2s^{2}-2p3d$} & 10.1512(6) & 10.175 & 700(20) & 10(2) & 290(60) & 2.7(2) & 16(2) & 5.3(7) \\
   
\multicolumn{3}{l}{Fe XXIII $2s^{2}-2p3s$} & --	 & 10.560 & --	& -- & 	-- & --	& -- & -- \\

\multicolumn{3}{l}{Fe XXIII $2s^{2}-2s3p$} & 10.9671(3) & 10.981 & 380(10) & 16(1) & 440(30) & 7.0(3) & 40(2) & 0.54(3) \\

\multicolumn{3}{l}{Fe XXIII $2s^{2}-2s3p4$} & 11.0049(5) & 11.018 & 360(10) & 17(1) & 460(30) & 6.8(3) & 33(2) & 1.1(1) \\

\multicolumn{3}{l}{Fe XXII $2s^{2}2p-2s^{2}3d$}	& 11.755(1) & 11.770 & 380(20) & 8(1) & 200(30) & 2.4(1) & 16.0(6) & 0.19(1) \\

\multicolumn{3}{l}{Fe XXII $2s^{2}2p-2s^{2}3d$} & 11.909(2) & 11.920 & 280(50) & 12(2) & 300(50) & 1.7(1) & 12.7(7) & 0.17(1) \\

\multicolumn{3}{l}{Ni XXVI $1s^{2}2s-1s^{2}6p$} & 6.045(1) & --	 & -- & 9.2(5) & 450(30) & 1.4(2) & 2.1(3) -- \\

\multicolumn{3}{l}{Ni XXVI $1s^{2}2s-1s^{2}5p$} & 6.103(1) & 6.120 & 830(50) & 16(2) & 800(100)	& 2.6(2) & 3.9(4) & 4.2(5) \\

\multicolumn{3}{l}{Ni XXVI $1s^{2}2s-1s^{2}4p$} & 6.8029(4) & 6.8163 & 650(20) & 27(1) & 1200(50) & 5.7(1) & 9.5(2) & 2.3(1) \\

\multicolumn{3}{l}{Ni XXVI $1s^{2}2s-1s^{2}3p$} & 9.0479(2) & 9.061 & 430(10) & 15.1(5)	& 500(20) & 6.6(1) & 18.5(3) & 1.04(1) \\

\multicolumn{3}{l}{Ni XXVI $1s^{2}2s-1s^{2}3p$} & 9.0917(4) & 9.105 & 440(10) & 11.8(4)	& 400(30) & 4.5(1) & 12.4(3) & 1.31(3) \\

\multicolumn{3}{l}{Co XXV $1s^{2}2s-1s^{2}3p$} & 9.782(1) & 9.795 & 400(40) & 8(4) & 200(100) & 1.0(2) & 3.2(6)	& 0.15(3) \\

\hline

\multicolumn{3}{l}{no ID; Al XIII $1s-5p$ ?} &	5.600(1) & -- & --&  6(2) & 400 & 0.85(15) & 1.1(2) & -- \\		

\multicolumn{3}{l}{no ID; Ni XXVI $2p-5d$ ?} & 6.250(1) & --  & -- & $<4$ & $<200$ & 1.0(2) & 1.5(3) & -- \\
 
\multicolumn{3}{l}{no ID; Ni XXVI $2p-4s$ ?} &	7.0555(8) & -- & -- & 7(2) & 300(100) & 1.3(1) & 2.4(2) & -- \\

\multicolumn{3}{l}{no ID; Ni XXVI $2p-4d$ ?} &	7.0815(8) & -- & -- & 12(2) & 500(100) & 3.0(3) & 5.4(5) & -- \\      

\multicolumn{3}{l}{no ID; Ni XXVI $2p-3d$ ?} &	9.3726(4) & -- & -- & 14(1) & 450(30) & 4.2(2) & 12.3(6) & -- \\

\multicolumn{3}{l}{no ID; Ni XXV $2s2p-2s3d$ ?} & 9.9509(5) & -- & -- & 6(2) & 200(50) & 2.3(2) & 7.7(6) & -- \\

\multicolumn{3}{l}{no ID} &                       6.8690(5) & -- & -- & 10(2) & 440(80) & 1.7(1) & 2.9 & -- \\

\multicolumn{3}{l}{no ID} &                       8.081(2) & -- & -- & 7(2) & 260(80) & 0.6(1)	& 1.4 & -- \\

\multicolumn{3}{l}{Fe XXII?} &                       8.960(1) & -- & -- & $<5$ & $<200$ & 1.0(1) & 2.6 & -- \\

\multicolumn{3}{l}{no ID} &                       10.1015(6) & -- & -- & 12(5) & 400(100) & 1.7(2) & 5.8 & -- \\

\multicolumn{3}{l}{no ID} &                       10.494(1) & -- & -- & 3(3) & 100(100) & 1.5(2) & 6.2 & -- \\

\multicolumn{3}{l}{no ID} &                       11.413(2) & -- & -- &	 57(9) & 1500(200) & 7(1) & 38.9 & -- \\ 

\multicolumn{3}{l}{Fe XXII?} &                       11.472(2) & -- & -- &	 12(5) & 300(100) & 1.6(3) & 10.7 & -- \\

\multicolumn{3}{l}{no ID} &                       12.593(2) &-- & -- & 19(7) & 500(200)	& 2.3(4) & 19.5 & -- \\

\multicolumn{3}{l}{no ID} &                       12.644(3) &-- & -- & 24(7) & 600(100)	& 2.1(4) & 18.7 & -- \\					

\hline
\end{tabular}
\vspace*{\baselineskip}~\\ \end{center} 
{Parameters of the disk wind absorption lines observed in
GRO~J1655$-$40 for elements with $Z\geq17$ are given above.  For
clarity, the lines are listed by element in order of ascending atomic
number, and in order of increasing wavelength by element and ion.  The
spectral continua were fit locally using power-law models modified by
neutral photoelectric absorption edges (due to the interstellar
medium) where appropriate.  The lines were fit with simple Gaussian
models.  The errors quoted above are $1\sigma$ uncertainties.  Line
significances were calculated by dividing line flux by its 1$\sigma$
error.  Where errors are not given, the parameter was fixed at the
quoted value.  Line widths consisent with zero are not resolved.}
\vspace{-1.0\baselineskip}
\end{scriptsize}
\end{table}

\clearpage

\centerline{~\psfig{file=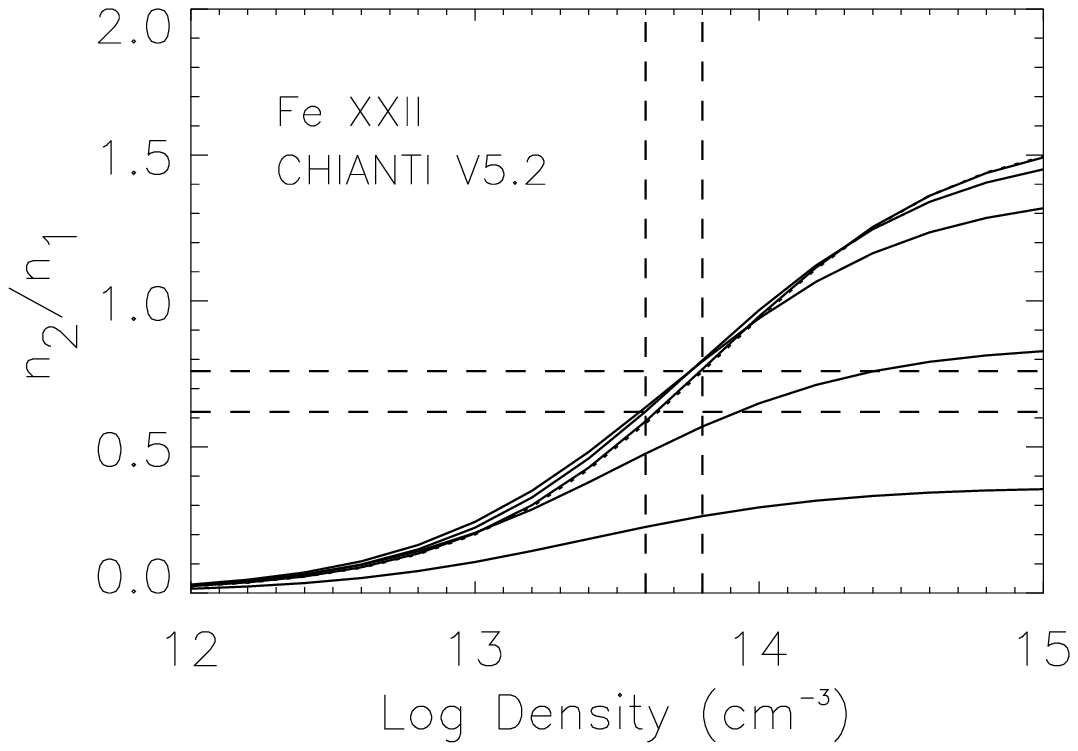,width=4.0in}~}
\figcaption[h]{\footnotesize The ratio of the populations of the fine
structure states of the ground level of Fe XXII predicted by CHIANTI
are shown here.  The curves correspond to temperatures log T = 5.0,
5.3, 5.7, 6.0 and 6.3 from bottom to top along the high density edge
of the plot.  The two Fe XXII lines detected in GRO~J1655$-$40
represent a density diagnostic that is as sensitive as better--known
He--like triplets.}
\medskip

\clearpage

\begin{table}[t]
\caption{Fits to the 1.3--2.5\AA~ Fe K Band with XSTAR and Cloudy Models}
\begin{footnotesize}
\begin{center}
\begin{tabular}{llllllllll}
Model & Code & Range & Profile & log($n_{0}$) & log($N_{H}$) & log($\xi$) & log($r_{0}$) & $\chi^{2}/\nu$ & $\chi^{2}/\nu$\\
\hline
1A & XSTAR kn & 1.3--2.5\AA & $r^{-2}$ & 13.0 & 23.68 & 3.0 & 10.7 & 6749/516 & 13.1 \\
1B & XSTAR ln & 1.3--2.5\AA & $r^0$ & 14.0 & $23.14(3)$ & 4.2 & 9.7 & 1670/517 & 3.23 \\ 
1C & XSTAR ln & 1.3--2.5\AA & $r^0$ & 14.0 & $23.1(1)$ & $4.80(5)$ & 9.44(3) & 1494/516 & 2.88 \\
\hline
2A & Cloudy & 1.3--2.5\AA & $r^{-2}$ & 13.0 & 23.75 & 3.0 & 10.7 & 2463/517 & 4.78 \\
2B & Cloudy & 1.3--2.5\AA & $r^0$ & 14.0 & 23.36(1) & 4.3 & 9.7 & 1232/517 & 2.43 \\ 
2C & Cloudy & 1.3--2.5\AA & $r^0$ & 14.0 & 23.70(1) & 5.7(2) & 9.0(1) & 1147/516 & 2.21 \\
\hline
\end{tabular}
\vspace*{\baselineskip}~\\ \end{center} 
{Fits to the 1.3--2.5\AA~ Fe K band with various XSTAR and CLOUDY
table models are listed above.  This region probes the hottest, most
ionized part of the wind, and therefore gives the best picture of the
innermost extent of the wind.  Models 1A and 2A were generated using
the thermal wind parameters given in Netzer (2006).  Models 1B and 2B
assume the same density as the best-fit models (1C and 2C), but fix
the radius at $r = 10^{9.7}$~cm, or $0.01\times R_{C}$.
Statistically, models 1B and 2B are vastly superior fits to the data
than models 1A and 2A.  However, the data and models are of such
quality that small changes in radius can be distinguished, and indeed
smaller radii are statistically required.  Models 1C and 2C are the
best-fit models; they are $7\sigma$ and $6\sigma$ improvements
over model 1B and 2B, respectively.  Models 2B and 2C have slightly
enhanced abundances of four times solar for Ca, Sc, Ti, V, Cr, and Mn
rather than merely twice solar as in the other models; these
abundances give slightly better fits.  For full details on all of
these models, please see the text.  Errors given in parentheses denote
the error in the last significant digit, and are $1\sigma$ confidence
errors.  Where errors are not given, the model parameter was fixed.}
\vspace{-1.0\baselineskip}
\end{footnotesize}
\end{table}

\begin{table}[t]
\caption{Broad-band fits with XSTAR and Cloudy Models}
\begin{footnotesize}
\begin{center}
\begin{tabular}{llllllllll}
Model & Code & Range & Profile & log($n_{0}$) & log($N_{H}$) & log($\xi$) & log($r_{0}$) & $\chi^{2}/\nu$ & $\chi^{2}/\nu$\\
\hline
\hline
3A & XSTAR kn & 1.3--13.3\AA & $r^{-2}$ & 13.0 & 23.68 & 3.0 & 10.7 & 340474/4794 & 71.0 \\
3B & XSTAR ln & 1.3--13.3\AA & $r^0$ & 14.0 & 23.46(1) & 4.3 & 9.7 & 47117/4793 & 9.83 \\ 
3C & XSTAR ln & 1.3--13.3\AA & $r^0$ & 14.0 & 23.65(1) & 4.88(2) & 9.41(2) & 43879/4792 & 9.16 \\
\hline
4A & Cloudy & 1.3--13.3\AA & $r^{-2}$ & 13.0 & 23.75 & 3.0 & 10.7 & 212478/4794 & 44.26 \\
4B & Cloudy & 1.3--13.3\AA & $r^0$ & 14.0 & $23.46(1)$ & $4.3$ & 9.7 & 87260/4793 & 18.20 \\ 
4C & Cloudy & 1.3--13.3\AA & $r^0$ & 14.0 & $23.67(2)$ & $4.88(2)$ & 9.40(2) & 85752/4792 & 17.89 \\
\hline

\end{tabular}
\vspace*{\baselineskip}~\\ \end{center} 
{Fits to the 1.3--13.3\AA~ band with various XSTAR and CLOUDY table
models are listed above.  Fitting single models to this broad bandpass
gives a measure of the average wind properties.  Models 3A and 4A were
generated using the thermal wind parameters given in Netzer (2006).
Models 3B and 4B assume the density used in the best-fit models (3C
and 4C), but fix the radius at $r = 10^{9.7}$~cm, or $0.01\times
R_{C}$.  Here again, models 3B and 4B are enormous statistical
improvements over models 3A and 4A, and smaller radii than those
assumed in models 3B and 4B are statistically required.  Models 3C and
4C are vast statistical improvements over models 3B and 4B.  Models 4B
and 4C have slightly enhanced abundances of four times solar for Ca,
Sc, Ti, V, Cr, and Mn rather than merely twice solar as in the other
models; these abundances give slightly better fits.  For full details
on all of these models, please see the text.  Errors given in
parentheses denote the error in the last significant digit, and are
$1\sigma$ confidence errors.  Where errors are not given, the model
parameter was fixed.}
\vspace{-1.0\baselineskip}
\end{footnotesize}
\end{table}

\clearpage

\centerline{~\psfig{file=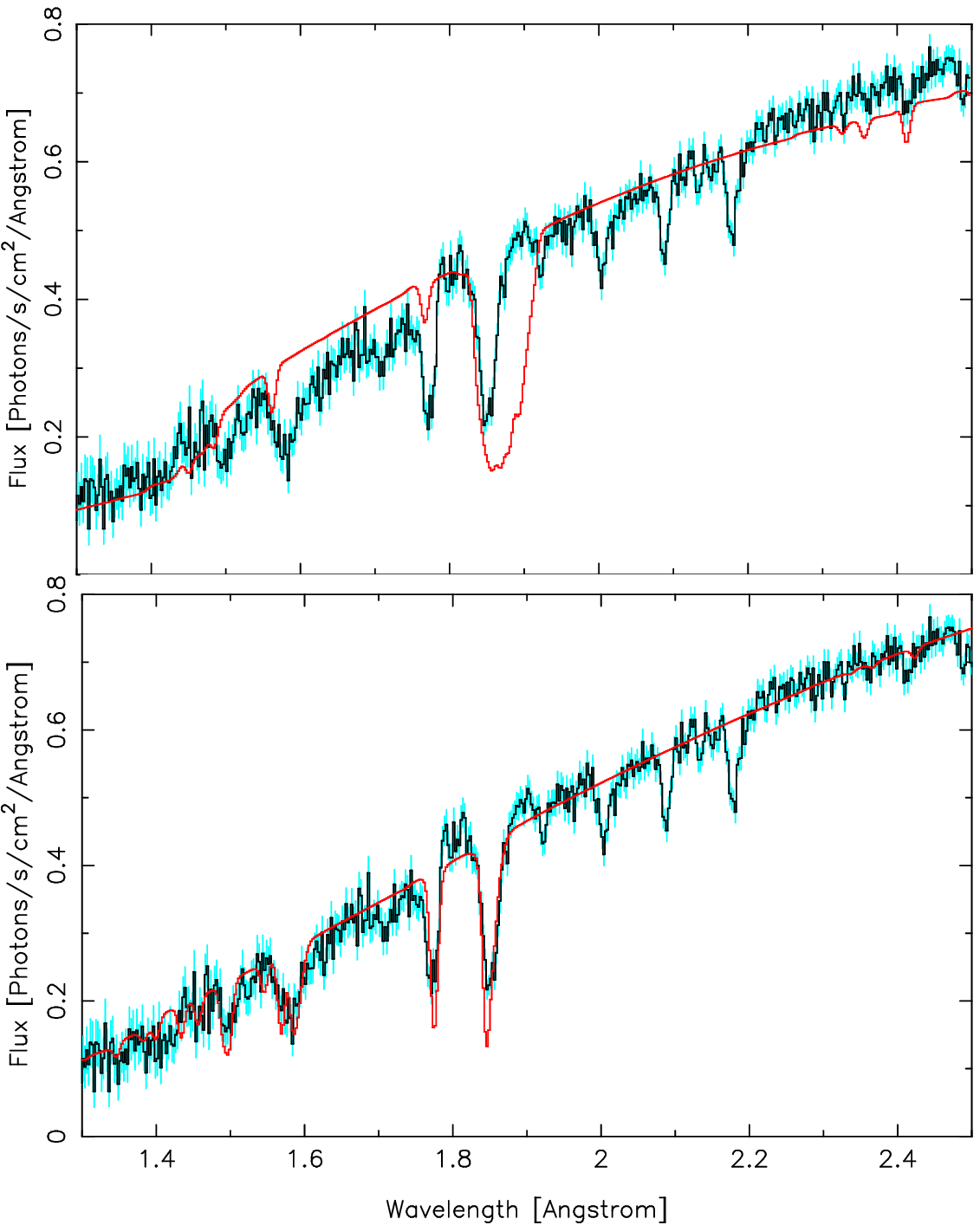,width=4.0in}~}
\figcaption[h]{\footnotesize The plot above shows the crucial Fe K
wavelength regime of the wind absorption spectrum, fit with XSTAR
models.  The He-like Fe XXV line at 1.850\AA~ and H-like Fe XXVI line
at 1.77\AA~ are expected to arise in the most ionized part of the
wind, and thus give the best measure of the innermost extent of the
wind.  Model 1A is shown in the upper panel; is consistent with a
thermally--driven wind, and is based on the models described in Netzer
(2006).  Model 1C is shown in the lower panel; it is the best-fit
XSTAR model, and corresponds to a wind where magnetic driving is
likely to be important.  For more details on the models shown above,
please see Table 3 and the text.}
\smallskip

\centerline{~\psfig{file=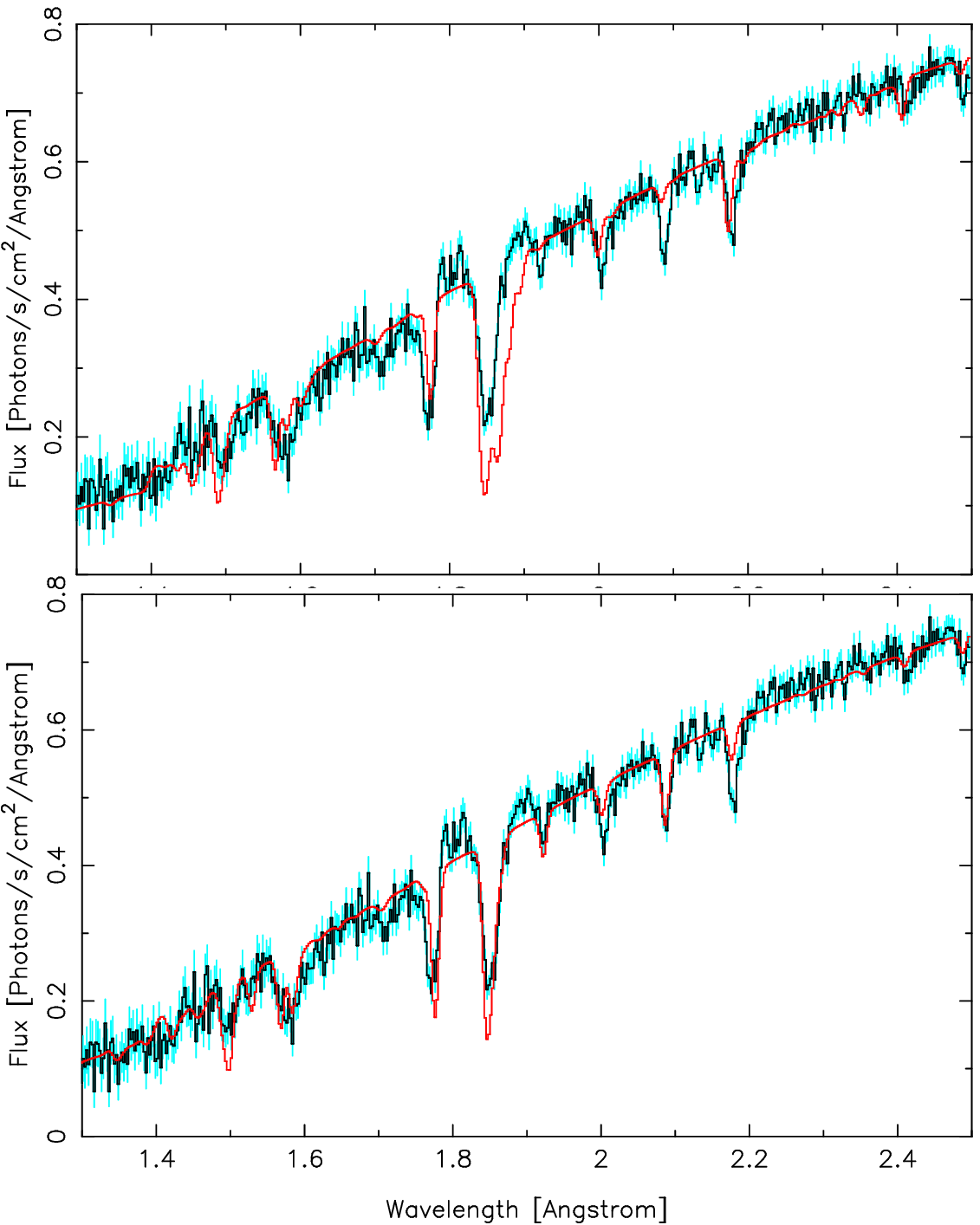,width=4.0in}~}
\figcaption[h]{\footnotesize The plot above again shows the crucial Fe
K wavelength regime of the wind absorption spectrum, fit with Cloudy
models.  The He-like Fe XXV line at 1.850\AA~ and H-like Fe XXVI line
at 1.77\AA~ are expected to arise in the most ionized part of the
wind, and thus give the best measure of the innermost extent of the
wind.  Model 2A is shown in the upper panel; is consistent with a
thermally--driven wind, and is based on the models described in Netzer
(2006).  Model 2C is shown in the lower panel; it is the best-fit
XSTAR model, and corresponds to a wind where magnetic driving is
likely to be important.  For more details on the models shown above,
please see Table 3 and the text.}
\medskip

\clearpage

\centerline{~\psfig{file=f7.ps,width=4.0in}~}
\figcaption[h]{\footnotesize The plot above shows the crucial short
wavelength regime of the rich wind absorption spectrum.  The model
shown above is the best-fit Cloudy model (2C), but is shown prior to
convolution with the HETGS response function.  The Fe XXV and Fe XXVI
absorption lines at 1.850\AA~ and 1.77\AA~ are actually black at line
center, though they do not appear to be black after convolving with
the instrument response (see the bottom panel in Figure 6).  To
address issues such as saturation, direct fitting is the most robust
means of modeling spectra.}
\medskip

\clearpage

\centerline{~\psfig{file=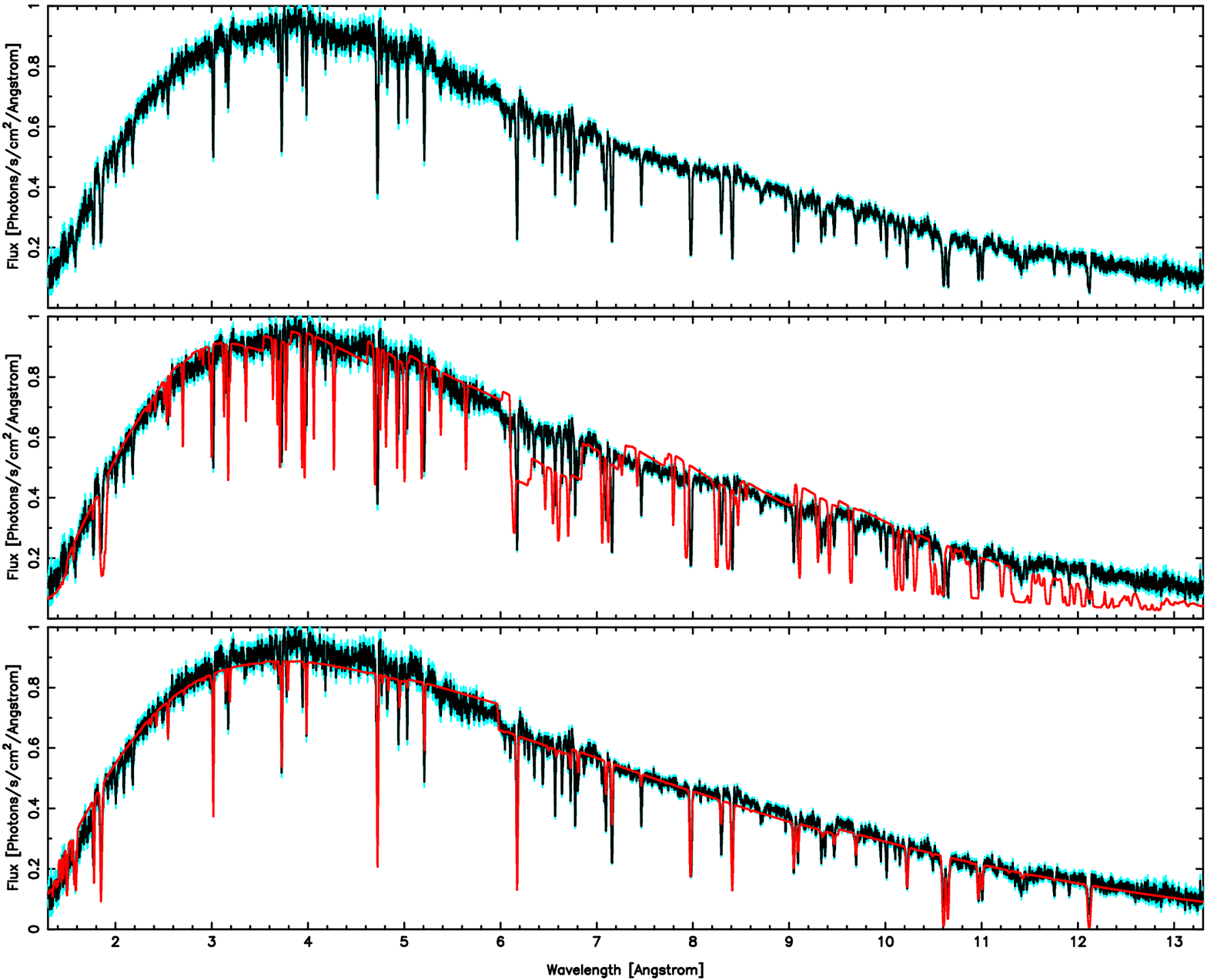,width=6.0in}~}
\figcaption[t]{\footnotesize The plot above shows the 1.3--13.3\AA~
spectrum of GRO J1655$-$40.  The top panel merely depicts the
spectrum.  The middle panel shows the spectrum fit with an XSTAR model
(3A) consistent with a thermally--driven wind, based on the models
described by Netzer (2006).  The bottom panel shows the spectrum fit
with our best-fit XSTAR model (3C), and corresponds to a scenario
where magnetic driving is likely to be important.  For details on the
models shown above, please see Table 4 and the text.}

\clearpage

\centerline{~\psfig{file=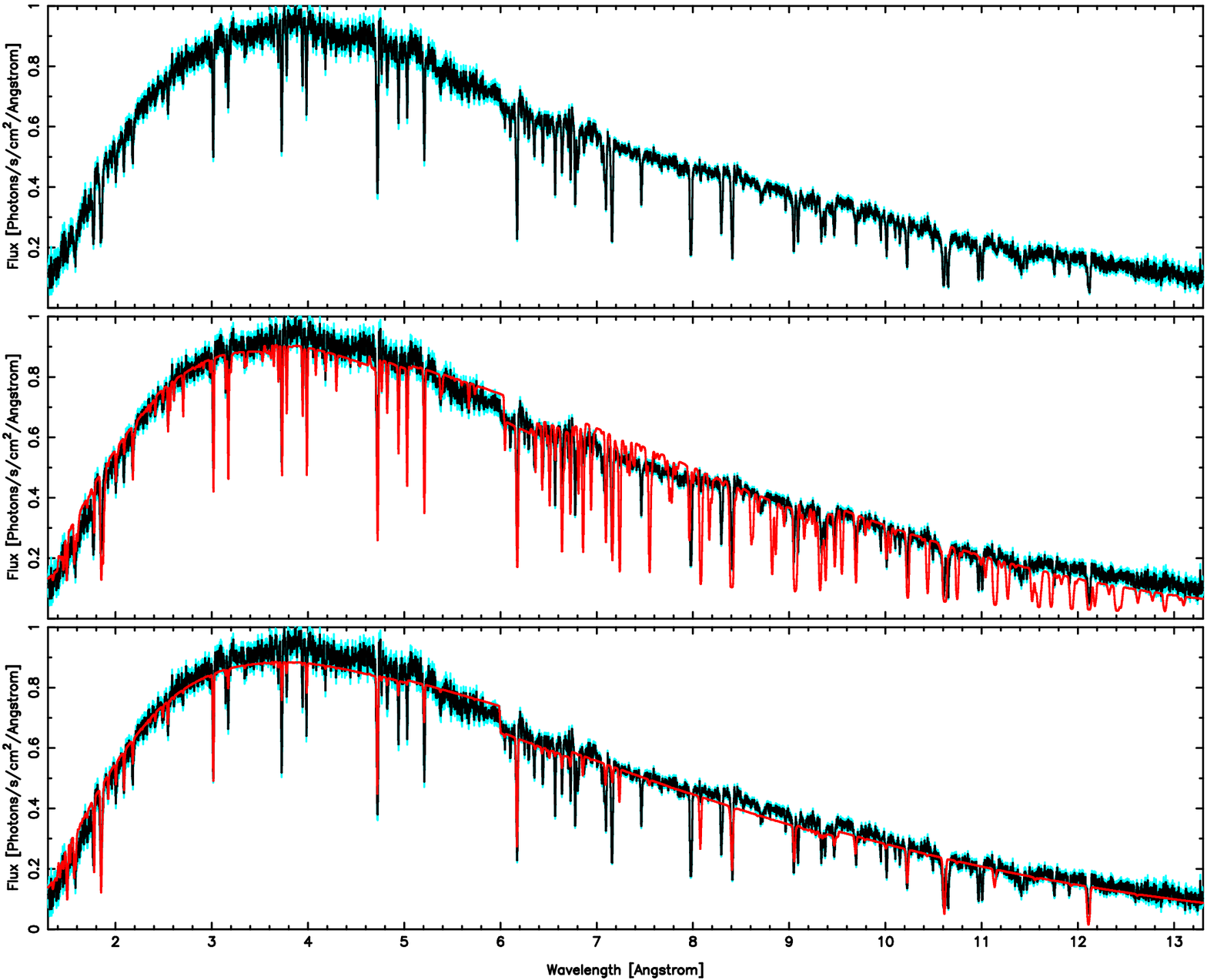,width=6.0in}~}
\figcaption[t]{\footnotesize The plot above shows the 1.3--13.3\AA~
spectrum of GRO J1655$-$40.  The top panel merely depicts the
spectrum. The middle panel shows the spectrum fit with a Cloudy model
(4A) consistent with a thermally--driven wind, based on the models
described by Netzer (2006).  The bottom panel shows the spectrum fit
with our best-fit Cloudy model, and corresponds to a scenario where
magnetic driving is likely to be important.  For details on the models
shown above, please see Table 4 and the text.}

\clearpage

\centerline{~\psfig{file=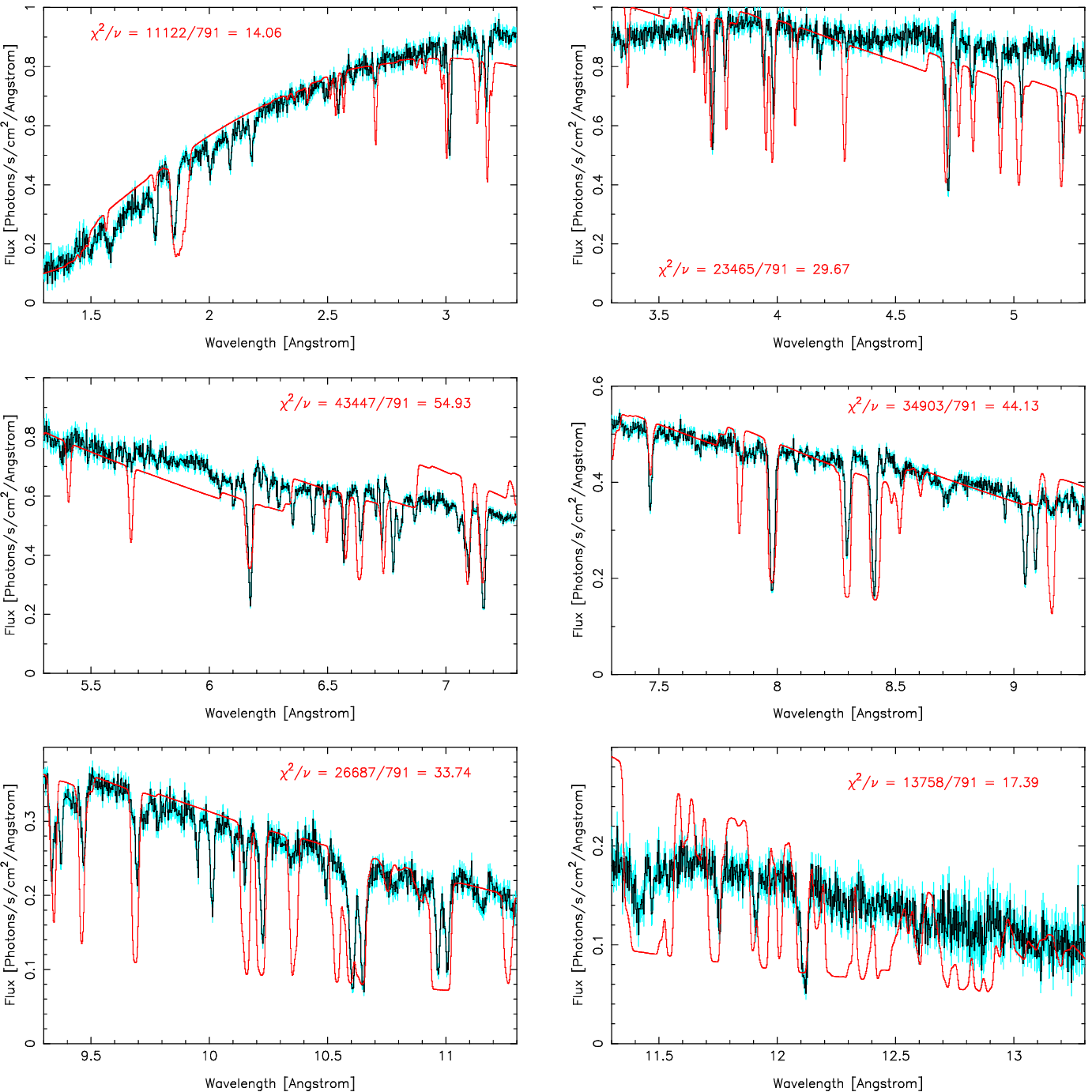,width=6.0in}~}
\figcaption[t]{\footnotesize The plot above shows the disk wind
spectrum of GRO J1655$-$40, fit in 2\AA~ slices with an XSTAR model
consistent with a thermally--driven wind (see models 1A and 3A in
Tables 3 and 4).  The model clearly fails to describe the spectrum in
each wavelength slice.  Apart from the ISM absorption and power-law
index (fixed to the best-fit values from broad-band fitting with {\it
RXTE}), the disk blackbody plus power-law continuum parameters were
allowed to vary.}

\clearpage

\centerline{~\psfig{file=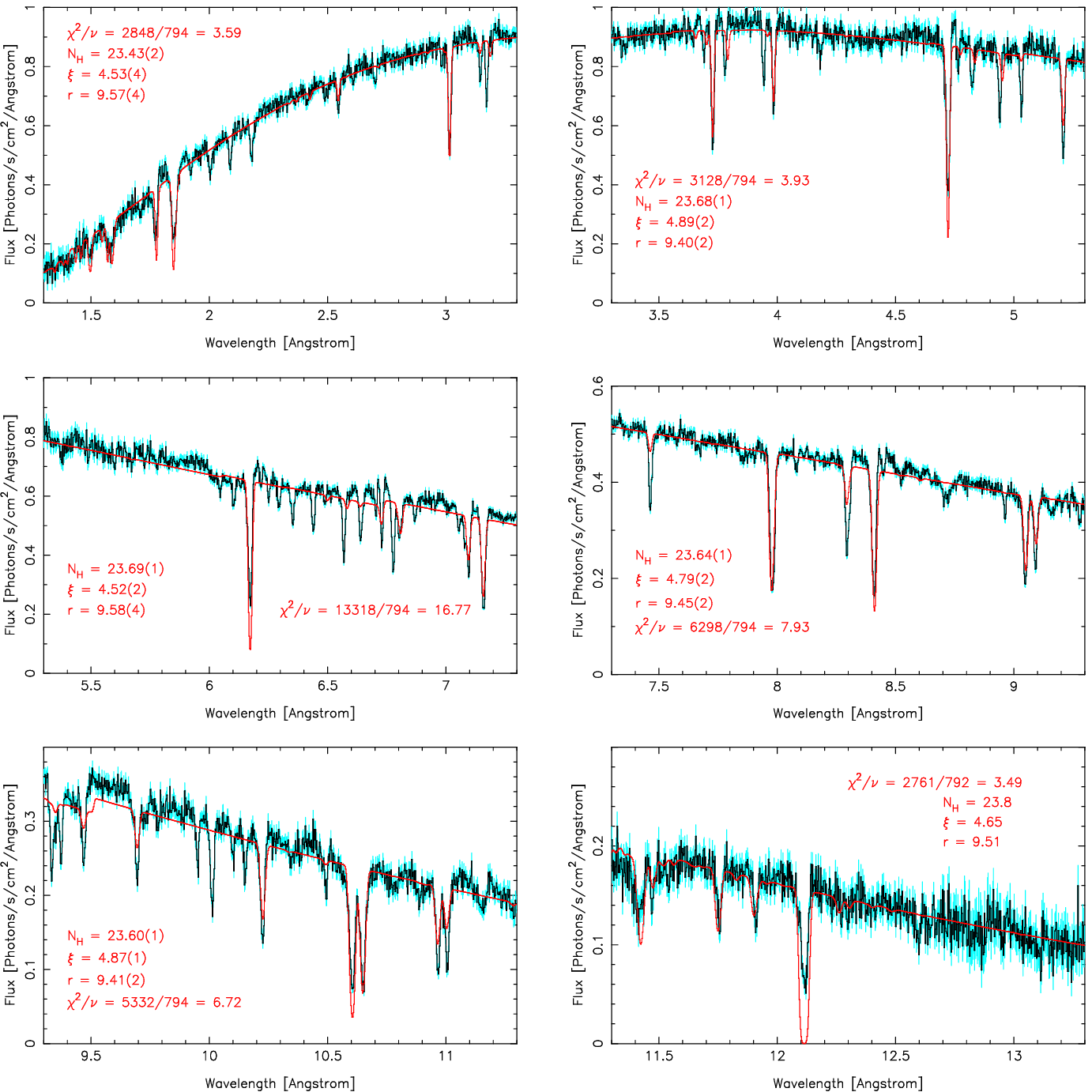,width=6.0in}~}
\figcaption[t]{\footnotesize The plot above shows the disk wind
spectrum of GRO J1655$-$40, fit in 2\AA~ slices with an XSTAR model
allowed to vary in each segment.  The absorption parameters derived
from each fit are given in the panel where the spectrum and fit are
plotted.  In all segments, the parameters are consistent with a wind
where magnetic driving is likely to be important.  The XSTAR model
depicted above is not a perfect description of the data, partially owing
to the range of velocities in each segment; however, it is
statistically far superior to the thermal wind model shown in Figure
10.}

\clearpage

\centerline{~\psfig{file=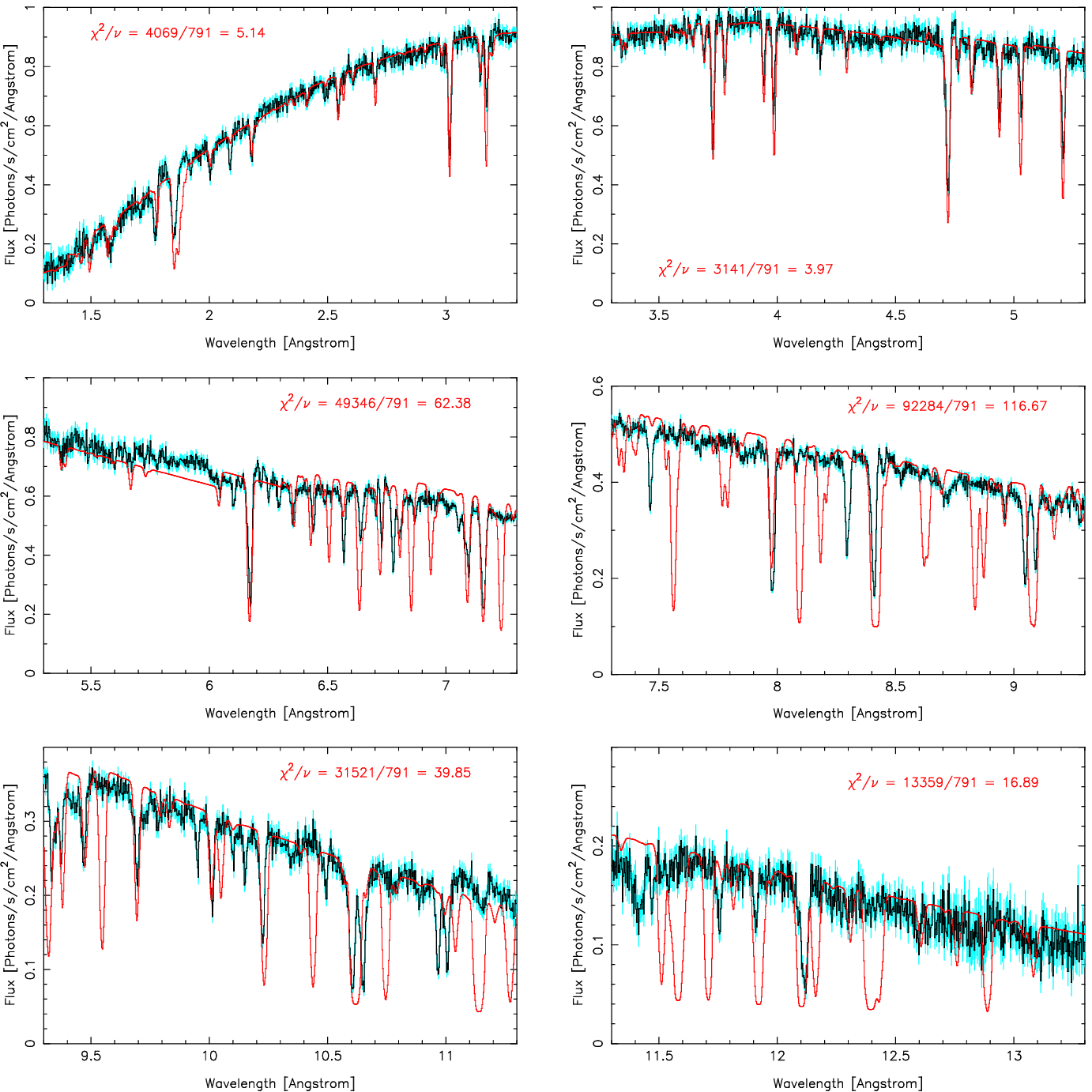,width=6.0in}~}
\figcaption[t]{\footnotesize The plot above shows the disk wind
spectrum of GRO J1655$-$40, fit in 2\AA~ slices with an Cloudy model
consistent with a thermally--driven wind (see models 2A and 4A in
Tables 3 and 4).  The model clearly fails to describe the spectrum in
each wavelength slice.  Apart from the ISM absorption and power-law
index (fixed to the best-fit values from broad-band fitting with {\it
RXTE}), the disk blackbody plus power-law continuum parameters were
allowed to vary.}
\clearpage

\centerline{~\psfig{file=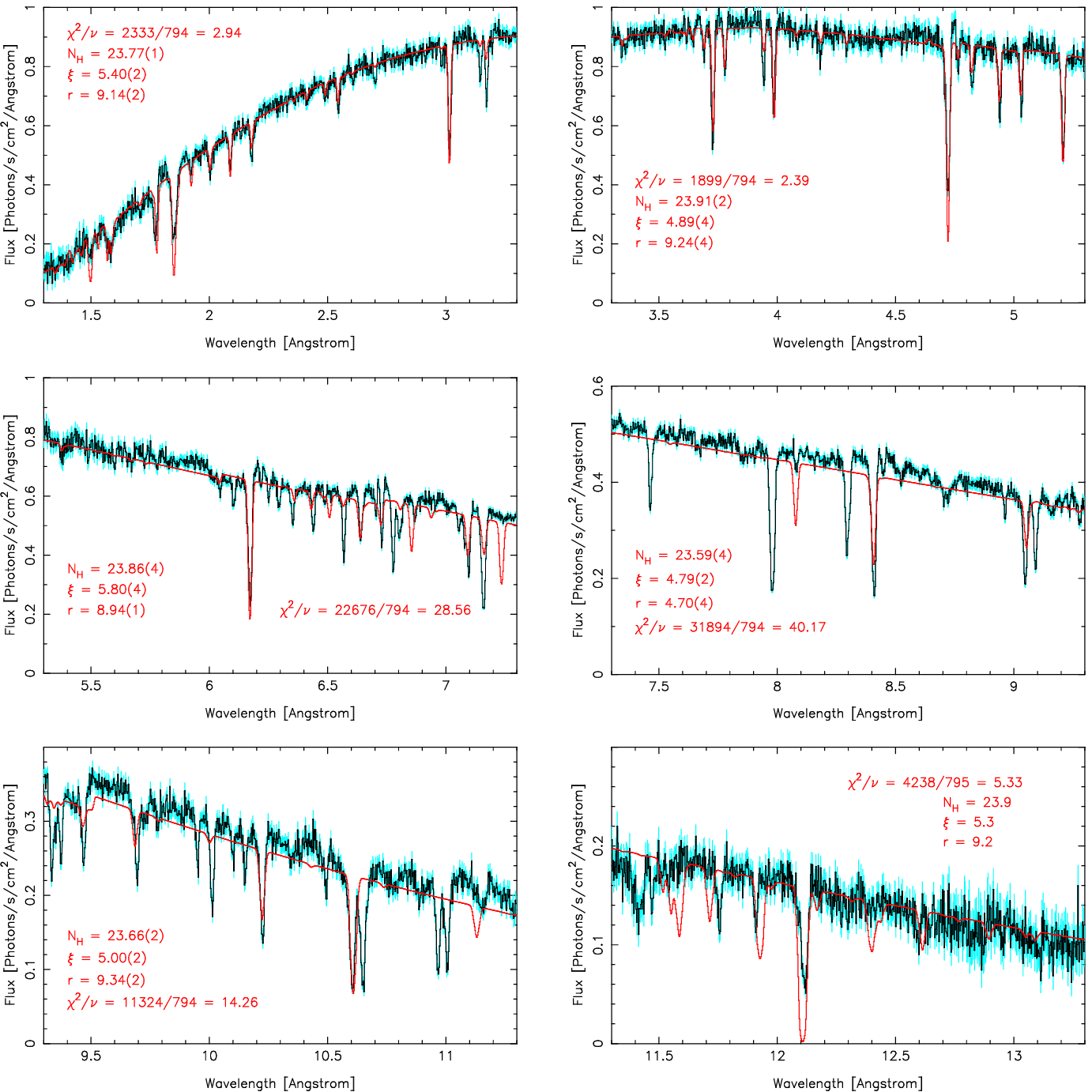,width=6.0in}~}
\figcaption[t]{\footnotesize The plot above shows the disk wind
spectrum of GRO J1655$-$40, fit in 2\AA~ slices with a Cloudy model
allowed to vary in each segment.  The absorption parameters derived
from each fit are given in the panel where the spectrum and fit are
plotted.  In all segments, the parameters are consistent with a wind
where magnetic driving is likely to be important.  The Cloudy model
depicted above is not a perfect description of the data, partially
owing to the range of velocities in each segment; however, it is
statistically far superior to the thermal wind model shown in Figure
12.}

\clearpage

\centerline{~\psfig{file=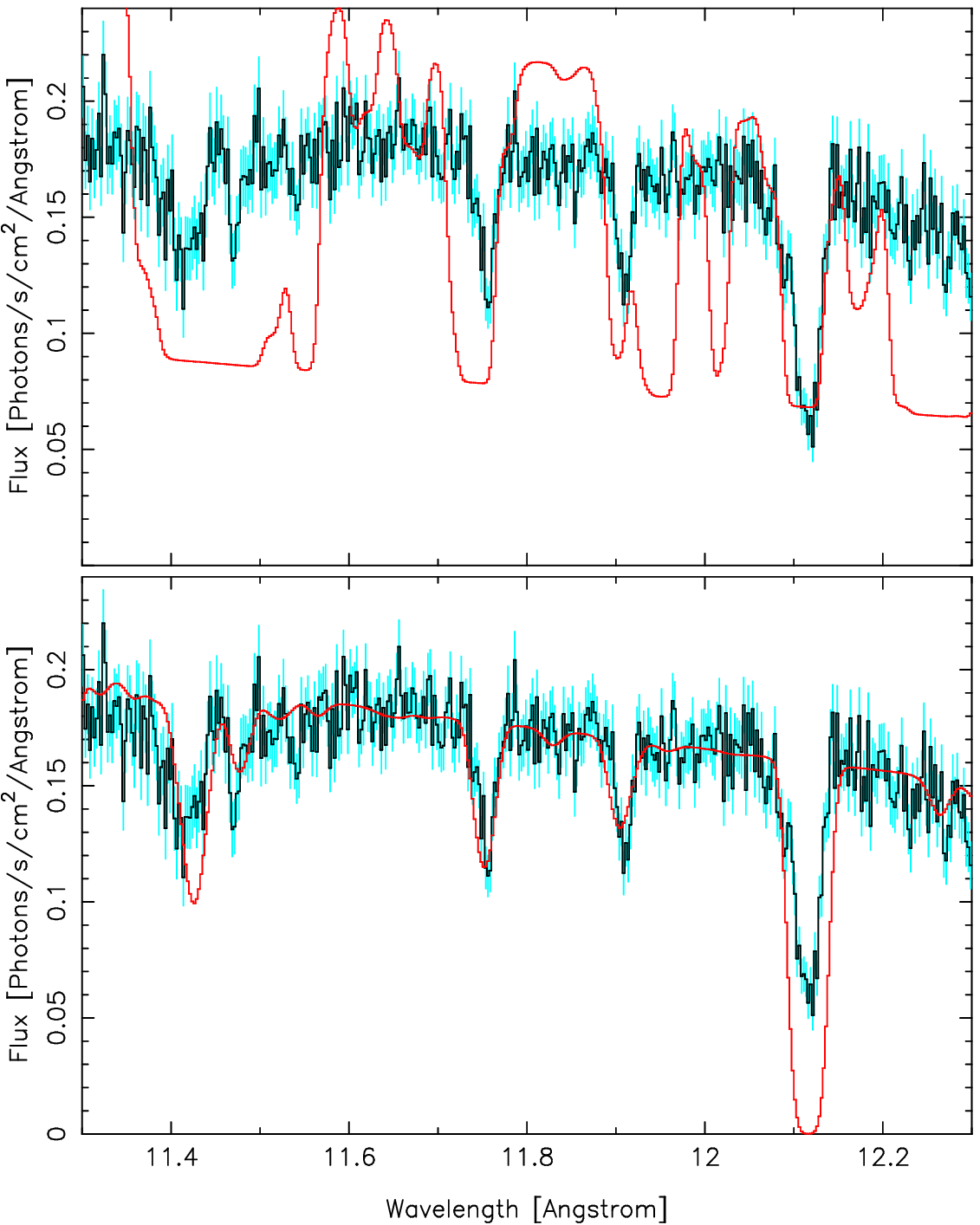,width=4.0in}~}
\figcaption[h]{\footnotesize The plots above highlight fits to the
density-sensitive Fe XXII lines at 11.77\AA~ and 11.92\AA.  The models
shown here are those depicted in Figures 10 and 11.  The upper panel
depicts a model consistent with a thermally--driven wind, based on the
parameters given in Netzer (2006).  The lower panel shows the best-fit
XSTAR model for this region; in this model, the lines are
well-represented and are clearly not saturated.  These models stongly
suggest that our estimate of the density (${\rm log(n)}
\simeq$13.7--14.0) is robust.  Cloudy models are not shown as the
wavelengths for these ions appear to be incorrect, but it is clear in
Figures 12 and 13 that our best-fit Cloudy models descibe the data far
better than alternative models.}
\smallskip

\end{document}